\documentclass[11pt, 
 reprint,
 notitlepage,
 amsmath,amssymb,
 aps,pra,superscriptaddress
]{revtex4-2}

\usepackage[T1]{fontenc}
\usepackage[colorlinks,citecolor=blue,urlcolor=blue]{hyperref} 
\usepackage{graphics}
\usepackage{subfigure}
\usepackage{natbib}
\usepackage{physics}
\usepackage{siunitx}
\usepackage{chemformula}

\usepackage{changes}
\definechangesauthor[name={Review},color=red]{Rev}

\begin{document}

\title{Classical Density Functional Theory Reveals Structural Information of \ch{H2} and \ch{CH4} Fluids Adsorbed in MOF-5}
\author{Elvis do A. Soares}%
\email{elvis.asoares@gmail.com}
\affiliation{Engenharia de Processos Químicos e Bioquímicos (EPQB), Escola de Química, Universidade Federal do Rio de Janeiro, 21941-909, Rio de Janeiro, RJ, Brazil}%
\author{Amaro G. Barreto Jr.}%
\affiliation{Engenharia de Processos Químicos e Bioquímicos (EPQB), Escola de Química, Universidade Federal do Rio de Janeiro, 21941-909, Rio de Janeiro, RJ, Brazil}%
\author{Frederico W. Tavares}%
\email{tavares@eq.ufrj.br}
\affiliation{Engenharia de Processos Químicos e Bioquímicos (EPQB), Escola de Química, Universidade Federal do Rio de Janeiro, 21941-909, Rio de Janeiro, RJ, Brazil}%
\affiliation{Programa de Engenharia Química, COPPE, Universidade Federal do Rio de Janeiro, 21941-909, Rio de Janeiro, RJ, Brazil}%

\date{July 5, 2023}

\begin{abstract}
    This study employs classical Density Functional Theory (cDFT) to investigate the adsorption isotherms and structural information of \ch{H2} and \ch{CH4} fluids inside MOF-5. The results indicate that the adsorption of both fluids is highly dependent on the fluid temperature and the shape of the MOF-5 structure. Specifically, the \ch{CH4} molecules exhibit stronger interactions with the MOF-5 framework, resulting in a greater adsorbed quantity compared to \ch{H2}. Additionally, the cDFT calculations reveal that the adsorption process is influenced by the fluid-fluid spatial correlations between the fluid molecules and the external potential produced by the MOF-5 solid atoms. These findings are supported by comparison with experimental data of adsorbed amount and the structure factor of the adsorbed fluid inside the MOF-5. We demonstrate the importance of choosing the appropriate grid size in calculating the adsorption isotherm and the fluid structure factors within the MOF-5. Overall, this work provides valuable insights into the adsorption mechanism of \ch{H2} and \ch{CH4} in MOF-5, emphasizing the importance of considering the structural properties of the adsorbed fluids in MOFs for predicting and designing their gas storage capacity at different thermodynamic conditions.
\end{abstract} 

\maketitle

\section{Introduction}

The emergent global energy demand has resulted in a pressing need for the efficient storage of fuel gases such as hydrogen (\ch{H2}) and methane (\ch{CH4}). In response to this challenge, one of the most promising solutions is the use of nanoporous materials for gas adsorption.~\cite{Broom2016,Anstine2022} These materials are characterized by their pore sizes that fall within the nanoscale range. To achieve effective gas storage, a detailed understanding of the adsorption processes of these gases on nanoporous materials is essential. 

Metal-organic frameworks (MOFs) are a class of porous materials that emerged as promising alternatives for gas storage~\cite{Li2018,Ding2019} and separation applications~\cite{Qian2020,Lin2020,Fan2021} due to their high surface areas~\cite{Qiu2014,Bavykina2020}, tailorable topologies, tunable functional groups, and chemical, mechanical and thermal stabilities \cite{Howarth2016,Yuan2018}. MOFs consist of metal nodes and organic linkers combined to form porous structures with a wide range of geometries, sizes, and functionalities, making them versatile materials for other applications, such as catalysis, sensing, and drug delivery. 

Some experimental studies have reported on the ability of MOFs to store \ch{H2}~\cite{Suh2012,Zhao2022,Kwon2022} and \ch{CH4}~\cite{Zhou2007,Kaye2007}, making them attractive materials for gas storage applications. For efficient gas storage, it is essential to model the adsorption processes of these gases on MOFs. This involves understanding the interaction between the gas molecules and the material surface, and the factors that affect adsorption, such as temperature and pressure. By modeling these processes, we can design MOFs with optimized properties for gas storage.
 
The traditional technique to study gas adsorption in MOFs or other materials is the Monte Carlo (MC) method~\cite{Keskin2009,Gallo2009,Getman2012,Prakash2013}. This method models the adsorption process at the molecular level, taking into account the complex interaction between the gas molecules and the MOF structure. As an example, the adsorption amount of \ch{H2}~\cite{Yang2006,Liu2009a,Fu2015} and \ch{CH4}~\cite{Duren2004,Pillai2015,Fu2015a,Altintas2016} in MOF-5 are well calculated from GCMC.

Another potential method to explore adsorption on nanoporous materials is the classical Density Functional Theory (cDFT) of fluids. In general, the formalism proposed by the cDFT allows evaluating the density distribution of the fluid inside the porous material, from which the absolute adsorbed amount can be easily obtained. Recently, some works have presented the ability of cDFT to calculate the adsorbed amount in MOFs materials. Liu \emph{et al.}~\cite{Liu2009a} utilized the weighted density approximation (WDA) for both repulsive and attractive contributions, and they calculated the \ch{H2} adsorption in MOF-5 and ZIF-8. They concluded that the WDA approach for the attractive contribution is more accurate than the Mean-Field Approximation (MFA). Fu \emph{et al.}~\cite{Fu2015a,Fu2015} used cDFT calculations to investigate \ch{H2} and \ch{CH4} storage capacity in several MOFs. They used the fundamental measure theory (FMT) to account for the repulsion effects, and they compared four versions of approximations to describe the attractive contribution: an MFA; a quadratic functional Taylor's expansion method; a full WDA; and an MFA corrected by WDA, the called novel weighted density functional (WDFT) proposed by Yu~\cite{Yu2009}. Their results demonstrate that WDFT is the most accurate formulation for MOF adsorption calculations, even at high pressures and low temperatures. Thereafter, FMT combined with WDFT has been utilized for high-throughput screening of MOFs for \ch{H2} storage~\cite{Liu2015}, \ch{CH4}/\ch{H2} separation~\cite{Guo2016}, and noble gas separation~\cite{Guo2018}. Trejos \emph{et al.}~\cite{Trejos2014,Trejos2018} have applied the SAFT-VRQ to model selectivities and adsorption isotherms of \ch{H2}/\ch{CH4} mixtures onto MOFs using an approximation of cDFT and 
semiclassical quantum correction. Very recently, Sang \emph{et al.}~\cite{,Sang2021} and Kessler \emph{et al.}~\cite{Kessler2021,Kessler2022} have combined the cDFT with PC-SAFT to calculate the adsorption of more complex fluids and mixtures in MOFs and COFs.

In this work, we use the cDFT of fluids to accurately model the intricate fluid-solid interactions and fluid-fluid correlations within MOF-5. Specifically, we investigate the behavior of \ch{H2} and \ch{CH4} fluids within the MOF-5 nanoporous material. According to the cDFT, the grand thermodynamic potential and Helmholtz free energy are functionals of the local density distribution, $\rho(\vb*{r})$. The interactions between the fluid particles can be taken into account by the use of the excess free energy functional. To achieve accurate predictions, we incorporate the FMT plus the WDFT as the excess free energy functional. The WDFT involves a corrected mean-field theory with a more accurate equation of state for the Lennard-Jones (LJ) fluid from Johnson, Zollweg, and Gubbins (JZG)~\cite{Johnson1993}, which improves the description of fluid-fluid spatial correlations and bulk thermodynamic properties. We investigate the grid size dependence of the cDFT calculation on the MOF-5 3D geometry. Our results show that the grid size slightly influences the \ch{H2} and \ch{CH4} adsorption isotherm calculation in MOF-5. We also discuss the ability of the cDFT to calculate the structure of the fluid inside the MOF-5 at different thermodynamic conditions. The grid size strongly influences the calculated structural information of the fluid inside the nanoporous material. Furthermore, we propose that the structure factor must be obtained experimentally or by GCMC simulation to be compared with our cDFT calculation in the future. Such information would be necessary to formulate new excess free energy functionals.

Following this section, Section~\ref{sec:theory} presents a brief background regarding the implementation of the cDFT and its application for describing gas adsorption in MOFs. The discussion of the results from the proposed strategy is performed in Section~\ref{sec:results} and compared with the experimental data of \ch{H2} and \ch{CH4} adsorption in MOF-5 from the literature. Finally, Section~\ref{sec:conclusions} presents the conclusions of this work.

\section{Theory and Methods}
\label{sec:theory}

\subsection{Classical Density Functional Theory} 

According to the cDFT of fluids~\cite{Henderson1992,Wu2008,Evans2009,Hansen2013}, the grand thermodynamic potential, $\Omega[\rho(\vb*{r})]$, and the Helmholtz free-energy, $F[\rho(\vb*{r})]$, are functionals of the local density distribution $\rho(\vb*{r})$. The grand potential functional $\Omega[\rho(\vb*{r})]$ is related to the free energy functional $F[\rho(\vb*{r})]$ by a thermodynamic relation given as
\begin{align}
    \Omega[\rho(\vb*{r})] = F[\rho(\vb*{r})] + \int_V \dd{\vb*{r}}[\phi_\text{ext}(\vb*{r}) - \mu]\rho(\vb*{r}),
    \label{eq:dft_grandpotential} 
\end{align}
where $\mu$ is the equilibrium chemical potential and $\phi_\text{ext}(\vb*{r})$ is an external potential acting on the fluid. The Helmholtz free energy functional is determined by the sum of two terms, in the form
\begin{align}
    F[\rho(\vb*{r})] = F_\text{id}[\rho(\vb*{r})] + F_\text{exc}[\rho(\vb*{r})] ,
    \label{eq:dft_freenergy}
\end{align}
where the first term is the ideal gas contribution and the second term is the excess free energy parcel (excess of ideal gas). The ideal-gas contribution $ F^\text{id}$ is given by the exact expression
\begin{align}
    F^\text{id}[\rho (\vb*{r})] = k_B T \int_{V} \dd{\vb*{r}} \rho(\vb*{r})[\ln(\rho (\vb*{r})\Lambda^3)-1],
\end{align}
where $k_B$ is the Boltzmann constant, $T$ is the absolute temperature, and $\Lambda$ is the well-known thermal de Broglie wavelength. The grand potential $\Omega[\rho(\vb*{r})]$ has  a minimum value when $\rho(\vb*{r})$ is the equilibrium density distribution, \emph{i.e.}, the minimum value of $\Omega[\rho(\vb*{r})]$ is the equilibrium grand potential of the system. Then, the equilibrium density profile should be calculated extremizing the grand canonical potential, such that
\begin{align}
    \left. \fdv{\Omega[\rho(\vb*{r})]}{\rho(\vb*{r})}\right|_{\mu,V,T} =\ & k_B T\ln(\rho (\vb*{r})\Lambda^3) + \fdv{F_\text{exc}[\rho(\vb*{r})]}{\rho(\vb*{r})} \nonumber \\
    & + \phi_\text{ext}(\vb*{r})- \mu= 0.
    \label{eq:dft_equilibrium_condition}
\end{align}
Using the definition of the chemical potential in the form $\mu = k_B T\ln(\rho_b\Lambda^3) + \mu_\text{exc}$, where $\rho_b$ is the uniform fluid bulk density, we can write the simplified form of Eq.~\eqref{eq:dft_equilibrium_condition} as 
\begin{align}
    \rho(\vb*{r}) = \rho_b \exp[-\beta \phi_\text{ext}(\vb*{r}) + \Delta c^{(1)}(\vb*{r}) ],
    \label{eq:dft_equilibrium_condition_simplified}
\end{align}
where $\beta = (k_B T)^{-1}$ is the inverse of the thermal energy, and the term $\Delta c^{(1)}(\vb*{r})$ is related to the first-order direct correlation function $c^{(1)}(\vb*{r})$ through the relation $\Delta c^{(1)}(\vb*{r}) = - \beta \fdv*{F_\text{exc}[\rho (\vb*{r})]}{\rho(\vb*{r})} + \beta \mu^\text{exc} $, which acts as a correction for the external potential $\phi_\text{ext}(\vb*{r})$ due to the fluid-fluid correlation.

The excess free energy functional $F_\text{exc}[\rho(\vb*{r})]$ contains all the information about the interaction between particles given by the pair potential $u(\vb*{r}-\vb*{r}')$. In our problem, the molecule-molecule interactions of the fluid can be well described by the Lennard-Jones potential in the form
\begin{align}
    u_\text{lj}(r) = 4\epsilon\left[ \left( \frac{\sigma}{r} \right)^{12}-\left( \frac{\sigma}{r} \right)^{6}\right].
\end{align}
To describe the $F_\text{exc}$ generated by both the repulsive short-range and the attractive long-range interactions of the LJ potential, we use the frequent separation of this excess free energy into repulsive and attractive contributions~\cite{Henderson1992,Peng2008,Lutsko2008d,Evans2009}, $F_\text{rep}$ and $F_\text{att}$, respectively, such that  
\begin{align}
    F_\text{exc}[\rho(\vb*{r})] = F_\text{rep}[\rho(\vb*{r})] + F_\text{att}[\rho(\vb*{r})].
    \label{eq:excess_free-energy}
\end{align}

The repulsive term $F_\text{rep}$ in Eq.~\eqref{eq:excess_free-energy} can be modeled by the free energy functional of a reference hard-sphere fluid~\cite{Kierlik1991} defined by a short-range hard-sphere potential given by
\begin{align}
    u_\text{hs}(r) = \begin{cases}
        \infty, \quad & r<d, \\
        0 , & r>d,
    \end{cases} 
\end{align}
where $d$ is an effective diameter of the hard-core repulsion. Barker and Henderson’s~\cite{Barker1967a,Barker1967} temperature-dependent diameter, defined by $d=\int_0^\sigma [1-e^{-\beta u_\text{lj}(r) }] \dd{r}$, can approximate this effective diameter. This integral can be approximated by the formula~\cite{Cotterman1986} 
\begin{align}
    \frac{d}{\sigma} =\frac{1+0.2977 T^*}{1+0.33163 T^* + 1.0477\times 10^{-3} T^{*2}},
\end{align}
where $T^* = k_B T/\epsilon$ is the reduced temperature. The modified fundamental measure theory (FMT)~\cite{Rosenfeld1989a} accurately describes the hard-sphere fluid structures and can represent the hard-sphere free energy functional, $F_\text{hs}[\rho(\vb*{r})]$. In this work, we have applied the antisymmetrized version of the White-Bear functional~\cite{Rosenfeld1997,Yu2002a,Roth2002} for the hard-sphere Helmholtz free energy contribution as
\begin{align}
    F_\text{rep}[\rho (\vb*{r})] = F_\text{hs}[\rho (\vb*{r})] = k_B T \int_{V} \dd{\vb*{r}} \Phi_\text{FMT}(\{ n^{(\alpha)}(\vb*{r})\}),
    \label{eq:hardsphere_free-energy}
\end{align}
where $\Phi_\text{FMT}$ is the local reduced free energy density of a mixture of hard-spheres, which is a function of the set of weighted densities, $n^{(\alpha)}(\vb*{r})$.

The attractive term $F_\text{att}$ in Eq.~\eqref{eq:excess_free-energy} represents the excess Helmholtz free energy contribution due to the particle-particle attractive interaction defined by the potential
\begin{align}
    u_\text{att}(r) = \begin{cases}
        0, \quad & r<\sigma, \\
        -\epsilon_1 \frac{e^{-\lambda_1(r/\sigma-1)}}{r/\sigma} -\epsilon_2 \frac{e^{-\lambda_2(r/\sigma-1)}}{r/\sigma},  & r>\sigma,
    \end{cases}
\end{align}
with $\epsilon_1 = -\epsilon_2 = 1.8577 \epsilon$, $\lambda_1 = 2.5449$, and $\lambda_2 = 15.4641$. The attractive parcel of the LJ potential was mapped onto a Two-Yukawa potential, as proposed previously~\cite{Kalyuzhnyi1996,Tang1997,Tang2001} to facilitate the Fourier Transform on the numerical calculations as discussed in Supporting Information. The free energy functional $F_\text{att}[\rho (\vb*{r})]$ can be described by the novel weighted density functional theory (WDFT)~\cite{Yu2009} as the sum of a mean-field term and a correlation contribution, respectively, in the form
\begin{align}
    F_\text{att}[\rho (\vb*{r})] =\ & \frac{1}{2} \int_{V} \dd{\vb*{r}} \int_{V} \dd{\vb*{r}'} \rho(\vb*{r}) u_\text{att}(|\vb*{r}-\vb*{r}'|) \rho(\vb*{r}') \nonumber \\
    &+ k_B T \int_{V} \dd{\vb*{r}} \Phi_\text{corr}(\bar{\rho}(\vb*{r})),
    \label{eq:wdft}
\end{align}
where the weighted density field here is given by $\bar{\rho}(\vb*{r})= \int_{V} \dd{\vb*{r}'} \rho(\vb*{r}')\omega_\text{wdft}(\vb*{r}-\vb*{r}')$ with $\omega_\text{wdft}(\vb*{r})= \Theta(d-|\vb*{r}|)/(4\pi d^3/3)$, and $\Theta(x)$ is the Heaviside function. The correlation free energy density is defined as  
\begin{align}
    \Phi_\text{corr}(\rho) = \beta \frac{F_\text{JZG}(\rho)}{V}-\beta \frac{F_\text{hs}(\rho)}{V}-\beta \rho^2 a_\text{mft},
\end{align}
giving the appropriated free energy for the bulk LJ fluid. The MFT parameter can be identified as the integral $a_\text{mft} = 2 \pi \int_0^\infty u_\text{att}(r) r^2 \dd{r} = -(16/9) \pi \epsilon \sigma^3$.

Finally, the absolute adsorption quantity can be calculated by the definition as 
\begin{align}
    N_\text{abs} = \int_{V _\text{uc}} \rho(\vb*{r}) \dd{\vb*{r}},
    \label{eq:adsorption_quantity}
\end{align}
where the local density distribution $\rho(\vb*{r})$ is given by Eq.~\eqref{eq:dft_equilibrium_condition_simplified}. The excess adsorption quantity, $N_\text{exc}$, is related to $N_\text{abs}$ by  
\begin{align}
    N_\text{exc} = N_\text{abs} - \rho_b V_\text{pore},
    \label{eq:excess_quantity}
\end{align}
where $\rho_b$ is the fluid bulk density, and $V_\text{pore}$ is the pore volume obtained from a helium (He) pycnometry~\cite{Myers2002}. This pore volume also can be calculated by the relation
\begin{align}
    V_\text{pore} = \int_{V _\text{uc}} \exp[-\beta \phi_\text{ext}^{(\text{He})}(\vb*{r})]\dd{\vb*{r}},
    \label{eq:helium_pycnometry}
\end{align}
where here $\phi_\text{ext}^{(\text{He})}(\vb*{r})$ is the fluid-solid interaction of a single He atom ($\sigma_\text{He} = 2.58~ \si{\angstrom}$, $\epsilon_\text{He}/k_B= 10.22~ \si{\kelvin}$)~\cite{Hirschfelder1964}. We can note that the exponential in Eq.~\eqref{eq:helium_pycnometry} is higher than unity at the regions where $\phi_\text{ext}^{(\text{He})}< 0$, and the exponential goes to zero at the solid atoms' positions. As Eq.~\eqref{eq:helium_pycnometry} is temperature-dependent, the pore volume is calculated at the standard temperature of 298 K.

The cDFT also possesses a notable capability of predicting the structural characteristics of the fluid confined within the porous material by utilizing the local density field, $\rho(\vb*{r})$. In our specific case, the local density field is a 3D field, which necessitates the utilization of isosurfaces to visualize our cDFT results. The structure factor $S(q)$ is related to the power spectral density by   
\begin{align}
    S (q) \propto |\tilde{\rho}_{\vb*{k}}|^2,
    \label{eq:structure_fator}
\end{align}
where $q= (k_x^2+k_y^2+k_z^2)^{1/2}$ is the magnitude of the wavenumber. Here, $\tilde{\rho}_{\vb*{k}}$ is the Fourier transform of the density profiles is computed as
\begin{align}
    \tilde{\rho}_{\vb*{k}} = \mathcal{F} \{ \rho(\vb*{r})\} = \int_V \dd{\vb*{r}} \rho(\vb*{r}) e^{-i\vb*{k}\cdot\vb*{r}}.
    \label{eq:ft}
\end{align}

\subsection{Equation of state for the fluid phase}

In this study, the Johnson, Zollweg, and Gubbins (JZG) equation of state for LJ fluids was employed~\cite{Johnson1993}. While this equation has been shown to be highly accurate, it is important to note that it is a semi-empirical relation with 33 parameters. This EoS is written as  
\begin{align}
    \frac{F_\text{JZG}}{V} = \rho \epsilon\sum_{i=1}^8 \frac{a_i (\rho \sigma^3)^i}{i} + \rho \epsilon \sum_{i=1}^6 b_i G_i,
\end{align}
where $a_i$ and $b_i$ are coefficients functions of temperature only. As reported in the original paper, the $G_i$ functions contain exponentials of the density and the one nonlinear parameter.
 
\begin{table}
    \caption{\label{tab:table_fluid}LJ parameters of the fluid molecules.}
    \centering
    \begin{ruledtabular}
    \begin{tabular}{cccc}
    Molecule & $\epsilon/k_B$ (K) & $\sigma\ (\si{\angstrom})$  & Model \\ \hline
    \ch{CH4} & 148.0 & 3.73 & TraPPE\footnotemark\\
    \ch{H2} & 34.20 & 2.96 & Buch\footnotemark \\
    \end{tabular}
    \end{ruledtabular} 
    \footnotetext[1]{Ref.~\cite{Martin1998}}
    \footnotetext[2]{Ref.~\cite{Buch1993}}
\end{table} 

\begin{figure}[htbp]
    \centering
    \includegraphics[width=\linewidth]{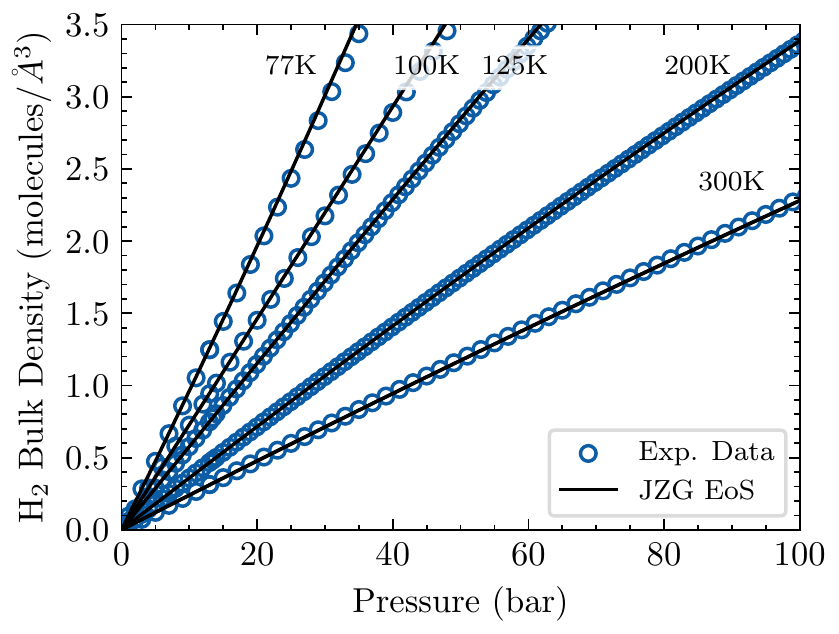}
    \caption{Equation of state of bulk \ch{H2} fluid over a broad pressure range and five different temperature values. Open symbols: experimental data~\cite{Leachman2009}. Solid lines: JZG EoS with LJ parameters from Table~\ref{tab:table_fluid}.}
    \label{fig:eos_H2}
\end{figure}

\begin{figure}[htbp]
    \centering
    \includegraphics[width=\linewidth]{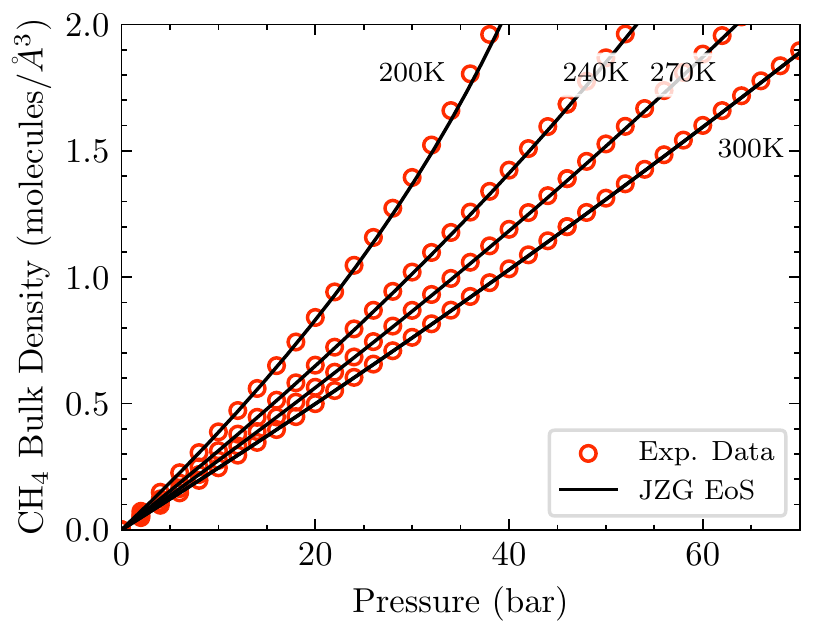}
    \caption{Equation of state of bulk \ch{CH4} fluid over a broad pressure range and four different temperature values. Open symbols: experimental data~\cite{Setzmann1991}. Solid lines: JZG EoS with LJ parameters from Table~\ref{tab:table_fluid}.}
    \label{fig:eos_CH4}
\end{figure}

The \ch{H2} and \ch{CH4} molecules are described with the LJ potential, which the parameters are given in Table~\ref{tab:table_fluid}. The \ch{H2} parameters were taken from the Buch work~\cite{Buch1993} without any quantum correction at low temperature. Other works~\cite{Darkrim1998,Wenzel2009,Fischer2009,Fischer2010} have demonstrated the influence of quantum corrections at low temperature and high pressure conditions, calculated by the Feynman-Hibbs perturbative approach~\cite{Feynman1965,Sese1994}, which modifies the LJ potential by adding a temperature-dependent term. The \ch{CH4} parameters are taken from the TraPPE force field~\cite{Martin1998}, as discussed in Supporting Information. The mapping of the equation of state is presented by the solid lines in Figure~\ref{fig:eos_H2} for \ch{H2} fluid and Figure~\ref{fig:eos_CH4} for \ch{CH4} fluid, where the open symbols are experimental data from Refs.~\cite{Leachman2009,Setzmann1991}.  

Although the LJ approximation for the fluid made of \ch{H2} molecules is sufficient at high temperatures, it is noted that this approximation presents slight deviations for temperatures below 100 K at higher pressure values, as shown in Figure~\ref{fig:eos_H2}.  However, making the quantum correction for \ch{H2} fluid at the temperature and pressure conditions presented here is unnecessary. Moreover, for the \ch{CH4} fluid, the LJ approximation is satisfactory even for the region below the critical temperature, as discussed in Supporting Information.

Also relevant for adsorption calculations, the molar mass of \ch{H2} is $m_{\ch{H2}} = 2.016$ g/mol and the molar mass of \ch{CH4} is $m_{\ch{CH4}} = 16.043$ g/mol. 

\subsection{External Potential produced by the solid atoms}

The total external potential produced by the solid atoms on the fluid molecules is represented as a sum of the Lennard-Jones interaction between the solid atoms and the fluid molecules as defined following
\begin{align}
    \phi_\text{ext}(\vb*{r}) = \sum_{i \in\ \text{solid}} u^{(\text{lj})}_{if}(|\vb*{r}-\vb*{r}_i|),
    \label{eq:external_potential}
\end{align}
with the mixed parameters obtained by the Lorentz-Berthelot mixing rules given by  $\sigma_{if} = \frac{1}{2}(\sigma_{ii}+\sigma_{ff})$ and $\epsilon_{if} = (\epsilon_{ii}\epsilon_{ff})^{1/2}$. The  LJ parameters for the MOF-5 atoms are taken from the DREIDING force-field~\cite{Mayo1990} and they are given in Table~\ref{tab:table_solid}.

\begin{table}
    \caption{\label{tab:table_solid}LJ parameters of the MOF-5 atoms.}
    \centering
    \begin{ruledtabular}
    \begin{tabular}{cccc}
    Model & Atom & $\epsilon/k_B$ (K) & $\sigma\ (\si{\angstrom})$ \\ \hline
    DREIDING\footnotemark  & H  & 7.649 & 2.846 \\
    & C  & 47.856 & 3.473  \\
    & O & 48.151  &  3.033  \\
    & Zn & 27.677 & 4.045 \\
    \end{tabular}
    \end{ruledtabular}
    \footnotetext[1]{Ref.~\cite{Mayo1990}}
\end{table}

To be consistent with the periodic boundary conditions imposed by the Fourier Transform, we replicate the solid in a 3x3x3 supercell with our unit cell in the center, where we calculate the external potential, Eq.~\eqref{eq:external_potential}, generated by the whole supercell. The CIF file from Ref.~\cite{Eddaoudi2002} was used to generate the MOF-5 structure in our calculations. 

\subsection{Numerical Methods}

To speed up the numerical calculations, we used of the Fast Fourier Transform (FFT) to calculate all the density convolutions. The analytical Fourier transform of the density profiles is computed from Eq.~\eqref{eq:ft}, where $\mathcal{F}$ represents the Fourier transform operator and $\mathcal{F}^{-1}$ represents the inverse Fourier transform operator given by 
\begin{align}
    \rho(\vb*{r}) \equiv \mathcal{F}^{-1}\{  \tilde{\rho}_{\vb*{k}}\}= \frac{1}{N_x N_y N_z}\sum_{\vb*{k}} \tilde{\rho}_{\vb*{k}}e^{i\vb*{k}\cdot\vb*{r}},
\end{align}
where $k_x = \{0,2\pi/L_x, 4\pi/L_x,\ldots,(N_x-1)2\pi/L_x\}$, $L_x$ is the length of the box in the $x$-direction, and $N_x=L_x/\Delta x$ is the number of gridpoints in the $x$-direction. And the same approach is used in other directions. We implemented the Fourier transform of the weight distribution $\tilde{\omega}_\alpha(\vb*{k})$ analytically in the reciprocal space. The weighted densities are then calculated 
\begin{align}
    n^{(\alpha)}(\vb*{r}) &= \int \dd{\vb*{r}'} \rho(\vb*{r}') \omega^{(\alpha)}(\vb*{r}-\vb*{r}') \nonumber \\
    &= \mathcal{F}^{-1} \{ \mathcal{F} \{ \rho(\vb*{r})\} \mathcal{F} \{ \omega^{(\alpha)}(\vb*{r})\}\}\nonumber \\
    &= \mathcal{F}^{-1} \{ \tilde{\rho}_{\vb*{k}} \tilde{\omega}^{(\alpha)}_{\vb*{k}}\}.
    \label{eq:convolution-weighteddensity}
\end{align}

All these convolutions, i.e., Eqs.~\eqref{eq:convolution-weighteddensity} were solved using the FFT functions from the \emph{Scipy} package.~\cite{2020SciPy} Our group implements, the FMT and WDFT functionals in \emph{Python} code.~\cite{Elvis2023} The Gibbs phenomenon was reduced by multiplicating the analytical Fourier transform $\tilde{\omega}_\alpha(\vb*{k})$ by the Lanczos $\sigma$-factor, $\sigma(k) = \sin(k/k_\text{max})/(k/k_\text{max})$, where $k_\text{max}$ is the maximum wavenumber from the FFT procedure.

The equilibrium condition for the cDFT, Eq.~\eqref{eq:dft_equilibrium_condition}, is solved using a fast inertial relaxation engine (FIRE) \cite{Bitzek2006,Guenole2020} also implemented in \emph{Python} by our group.~\cite{Elvis2020,Sermoud2021} The FIRE parameteres used in this work are $\alpha =0.2$ and $dt = 0.01$. The algorithm convergence criterion is 
\begin{align}
    \frac{1}{\sqrt{N_x N_y N_z}}\norm{\frac{\beta \fdv*{\Omega}{\rho(\vb*{r})}}{(atol + rtol|\rho(\vb*{r})|)} }_F \leq 1,
\end{align}
 with $atol = 10^{-6}$ and $rtol = 10^{-4}$, where $\norm{A}_F$ is the Frobenius norm. The initial density was considered uniform and equal to $\rho_b$. In the highly repulsive region, where $\phi_\text{ext}(\vb*{r})/k_B > 1.6 \times 10^4$ K, the initial density $\rho(\vb*{r})$ was assumed to be zero. For the adsorption isotherm calculations, the density profile obtained at low pressures is used as the initial profile of the next step with higher pressure. We have used a pressure step of 0.1 bar from 0.1 bar to 2.0 bar, a step of 1 bar from 2 bar to 10 bar, and a step of 10 bar from 20 bar to 500 bar.

\begin{figure}[htbp]
    \centering
    \includegraphics[width=\linewidth]{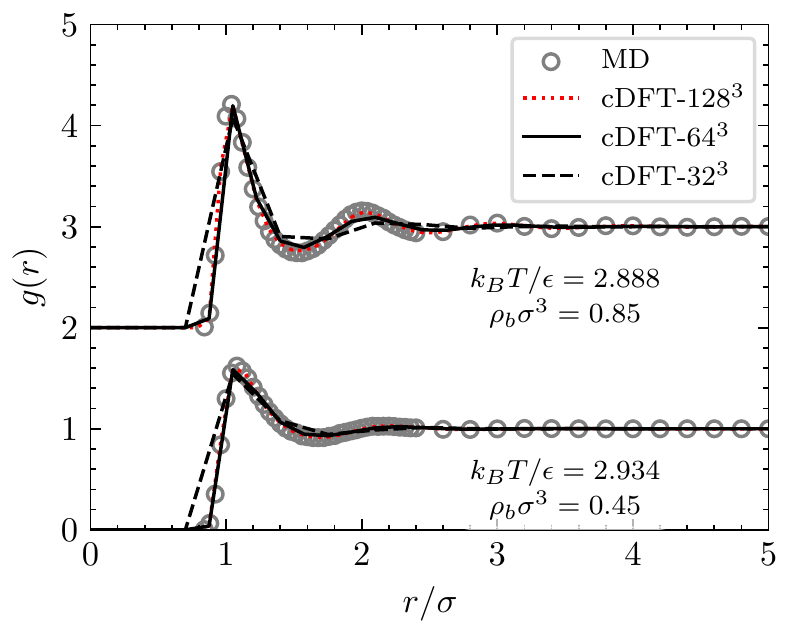}
    \caption{Radial distribution function of LJ fluid for two different densities and temperatures at supercritical conditions. Open symbols: MD data.~\cite{Reatto1986} Lines: our cDFT-3D results with different grid sizes. The curves for $\rho_b \sigma^3=0.85$ are shifted by 2 units in the $y$-axis.}
    \label{fig:rdf}
\end{figure} 

Figure~\ref{fig:rdf} illustrates an example of the Radial Distribution Function (RDF) calculation using our algorithm. The computation of the RDF, $g(r)$, can be effectively conducted utilizing the Percus test-particle approach. This method hinges on the principle that the system maintains invariance when a particle is stabilized at the origin; the pair correlation function is equivalent to the reduced density profile of the fluid around the fixed particle. We made the calculations using a box with a length of $L = 11.2\sigma$ such that the grid size is $\Delta x = 0.35 \sigma$ ($N^3=32^3$), $\Delta x = 0.175 \sigma$ ($N^3=64^3$), and $\Delta x = 0.0875\sigma$ ($N^3=128^3$). An adequate number of grid points is required to reproduce the fluid's structural information of the RDF compared to the MD simulation results. For temperatures above the critical point, a coarser grid size ($\Delta x = 0.35 \sigma$) can be used to make the cDFT calculations. Still, it presents small variations in the density profile, as presented in Fig.~\ref{fig:rdf}. The finest grid size ($\Delta x = 0.0875\sigma$) generates almost the same results as the medium grid size ($\Delta x = 0.175 \sigma$). However, this medium grid size has the advantage of lower time consumption to calculate.

Our cDFT calculations are made using the number of grid points $N^3=32^3$ and $N^3=64^3$. For the cases studied here, the results with $N^3=128^3$ are identical to the results with $N^3=64^3$. As the MOF-5 has a lattice length of $L = 25.866\ \si{\angstrom}$ with the unit cell volume of $V _\text{uc} = L^3 = 17305.325\ \si{\angstrom^3}$, the corresponding grid sizes are, respectively, $\Delta x \approx 0.81~ \si{\angstrom}$ ($N^3=32^3$) and $\Delta x \approx 0.40~ \si{\angstrom}$ ($N^3=64^3$). These grid sizes are related to the molecule diameter as $\Delta x \approx 0.27\sigma$ ($N^3=32^3$) and $\Delta x \approx 0.14\sigma$ ($N^3=64^3$) for the \ch{H2} molecule, and $\Delta x \approx 0.22\sigma$ ($N^3=32^3$) and $\Delta x \approx 0.1\sigma$ ($N^3=64^3$) for the \ch{CH4} molecule. Our systematic evaluations indicate that a grid size below $0.25\sigma$ is most suitable for the cDFT calculations. For comparison, a grid size of 0.59 \si{\angstrom} was used in Ref.~\cite{Liu2009a} and 0.5 \si{\angstrom} in Ref.~\cite{Fu2015,Fu2015a}. 

As an example of grid size dependence, we made calculations of the pore volume using Eq.~\eqref{eq:helium_pycnometry} as an Helium pycnometry. The Table~\ref{tab:table_helium} presents our calculated void fraction of MOF-5 from \ch{He} pycnometry, $\text{\ch{He} void fraction} = V_\text{pore}/V_\text{uc}$. This result demonstrates that there is a convergence of the results from $N^3 = 64^3$.

\begin{table}
    \caption{\label{tab:table_helium}Grid size dependence of the Helium void fraction.}
    \centering
    \begin{tabular}{cc}
        \hline \hline
    $N^3$ & He void fraction \\ \hline
    $32^3$ & 0.804 \\
    $64^3$ & 0.806 \\
    $128^3$ & 0.806 \\
    \hline \hline
    \end{tabular}
\end{table}

\section{Results and Discussion}
\label{sec:results}

\begin{figure}[htbp]
    \centering
    \includegraphics[width=\linewidth]{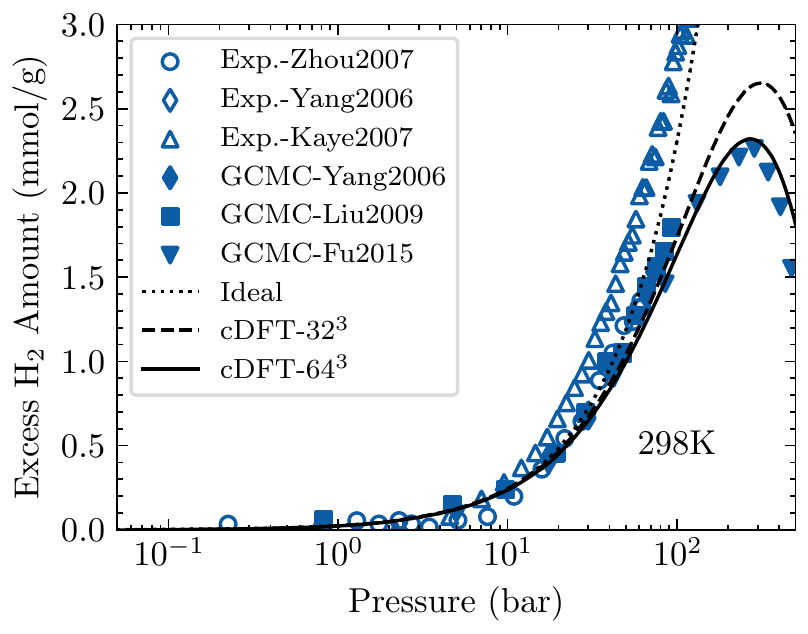}
    \caption{Excess amount of \ch{H2} in MOF-5 at 298 K over a wide pressure range. Open symbols: experimental data~\cite{Zhou2007,Yang2006,Kaye2007}. Closed Symbols: GCMC simulation data~\cite{Yang2006,Liu2009a,Fu2015}. Dotted lines: Ideal gas approximation. Lines: our cDFT results with different number of grid points.}
    \label{fig:H2_excess_adsorption_MOF5}
\end{figure}

Figure~\ref{fig:H2_excess_adsorption_MOF5} presents our calculated cDFT results of the excess amount of \ch{H2} in MOF-5 at 298 K over from 0.5 bar to 500 bar in a logarithm scale. We compare our results with experimental data~\cite{Zhou2007,Yang2006,Kaye2007} and GCMC simulation data~\cite{Yang2006,Liu2009a,Fu2015}. As discussed before, the cDFT calculations with $N^3=64^3$ tend to perform better than the cDFT calculations using $N^3=32^3$ mainly at high pressure region when compared to the GCMC simulation data. Beyond that, our cDFT results and GCMC data from Fu \emph{et al.}~\cite{Fu2015} predict a peak of excess amount of \ch{H2} in MOF-5 at 298 K around a pressure of 300 bar. On the other hand, the ideal gas approximation represented by the dotted line in Fig.~\ref{fig:H2_excess_adsorption_MOF5}, with the local density given by $\rho(\vb*{r}) = \rho_b \exp(-\beta \phi_\text{ext}(\vb*{r}))$, cannot describe this behavior of the \ch{H2} fluid inside the MOF-5 at pressures above 10 bar. We found considerable difference between the experimental data from Kaye~\cite{Kaye2007} and the other experimental data and GCMC data.

\begin{figure}[htbp]
    \centering
    \includegraphics[width=\linewidth]{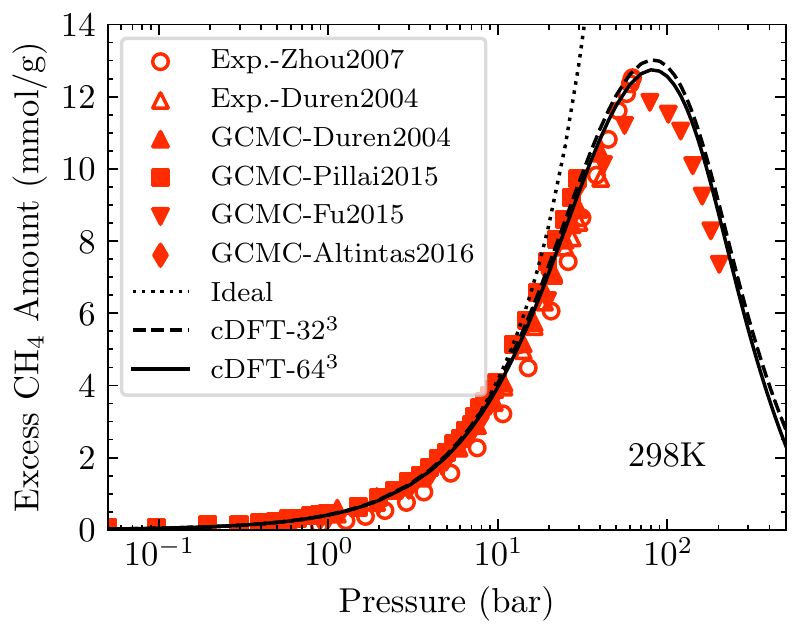}
    \caption{Excess amount of \ch{CH4} in MOF-5 at 298 K over a wide pressure range. Open symbols: experimental data~\cite{Zhou2007,Duren2004}. Closed Symbols: GCMC simulation data~\cite{Duren2004,Pillai2015,Fu2015a,Altintas2016}. Dotted lines: Ideal gas approximation. Lines: our cDFT results with different number of grid points.}
    \label{fig:CH4_excess_adsorption_MOF5} 
\end{figure}

Similarly, Figure~\ref{fig:CH4_excess_adsorption_MOF5}  presents our calculated cDFT results of the excess amount of \ch{CH4} in MOF-5 at 298 K over a wide pressure range in a logarithm scale. We also compare our cDFT results with experimental data~\cite{Zhou2007,Duren2004} and GCMC simulation data~\cite{Duren2004,Pillai2015,Fu2015a,Altintas2016}. Here, the cDFT calculations with $N^3=64^3$ and $N^3=32^3$ are similar when compared to the GCMC simulation data. Again, our cDFT results and GCMC data from Fu \emph{et al.}~\cite{Fu2015a} predict a peak of excess amount of \ch{CH4} in MOF-5 at 298 K around 80 bar. The experimental excess amount of \ch{CH4} in MOF-5 reported in Ref.~\cite{Zhou2007} is expected to increase after 60 bar although our cDFT results and the GCMC results from Ref.~\cite{Fu2015a} present a different behavior. The ideal gas approximation cannot describe the behavior of the excess adsorption of the \ch{CH4} fluid inside the MOF-5 at pressures above 10 bar. Thereby, the \ch{CH4} and \ch{H2} excess adsorption amounts are dominated by the fluid-fluid correlations at pressure above 10 bar. These results demonstrate the importance of taking into account fluid-fluid correlation effects in adsorption calculations in nanoporous materials. 

The total gravimetric uptake, $N_a$ (in wt\%), is usually reported in experimental measurements~\cite{Ahmed2017}. It can be calculated from the absolute adsorption quantity, $N_\text{abs}$ from Eq.~\eqref{eq:adsorption_quantity}, by the relation
\begin{align}
    N_a = \frac{m_f N_\text{abs}}{\rho_\text{cr}V _\text{uc} +m_f N_\text{abs}} \times 100,
\end{align}
where $m_f$ is the fluid molecule mass, $\rho_\text{cr} V _\text{uc}$ is the mass of the MOF unit cell, and the crystal density is $\rho_\text{cr} = 0.593\ \si{g/\centi\meter^3}$. 

\begin{figure}[htbp]
    \centering
    \includegraphics[width=\linewidth]{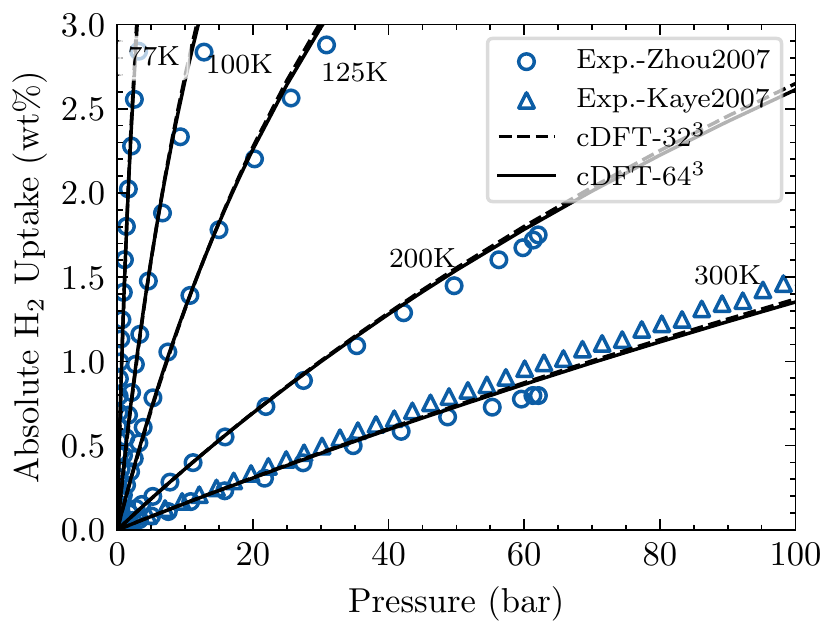}
    \caption{Absolute adsorption isotherms of \ch{H2} in MOF-5 at temperature values of 77 K, 100 K, 125 K, 200 K and 300 K and a wide pressure range. Open symbols: experimental data~\cite{Zhou2007,Kaye2007}. Solid lines: our cDFT results.}
    \label{fig:H2_absolute_adsorption_MOF5}
\end{figure}

Fig.~\ref{fig:H2_absolute_adsorption_MOF5} shows the absolute isotherms of \ch{H2} adsorption in MOF-5 at 77 K, 100 K, 125 K, 200 K, and 300 K in a wide range of pressure values. The cDFT predictions align well with the experimental data from Ref.~\cite{Zhou2007,Kaye2007}. These isotherms demonstrate that the amount of \ch{H2} adsorbed is significantly impacted by temperature. This can be attributed to the local density dependence on the external potential and temperature in Eq.~\eqref{eq:dft_equilibrium_condition_simplified}. As a result, when the temperature increases, the effect of the external potential on the local density decreases exponentially.  

\begin{figure}[htbp]
    \centering
    \includegraphics[width=\linewidth]{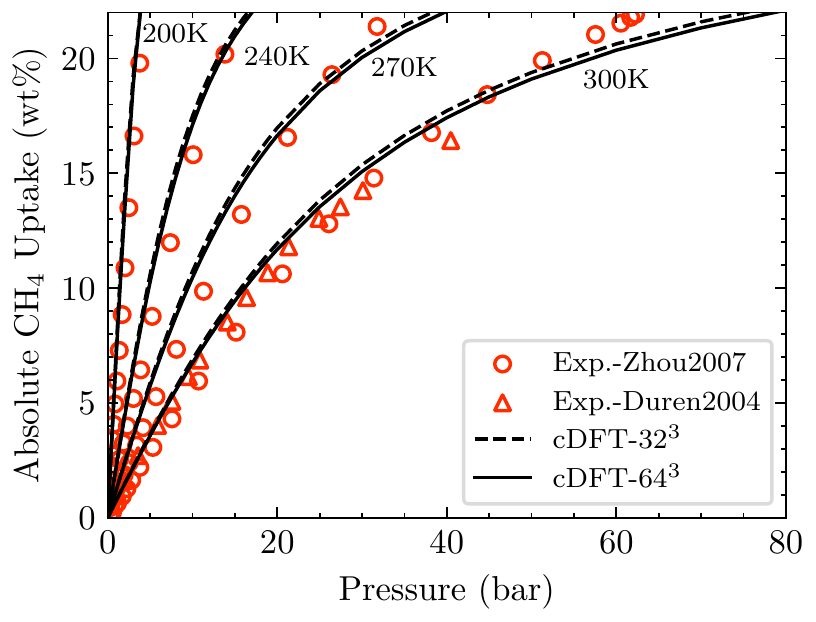}
    \caption{Absolute adsorption isotherms of \ch{CH4} in MOF-5 at temperature values of 200 K, 240 K, 270 K and 300 K and a wide pressure range. Open symbols: experimental data~\cite{Zhou2007,Duren2004}. Solid lines: our cDFT results.}
    \label{fig:CH4_absolute_adsorption_MOF5}
\end{figure}

Similarly, Figure~\ref{fig:CH4_absolute_adsorption_MOF5} shows the absolute isotherms of \ch{CH4} adsorption in MOF-5 at 200 K, 240 K, 270 K, and 300 K and the same pressure range. Overall,  the \ch{CH4} isotherms demonstrate higher adsorbed amount values than the \ch{H2} isotherms in MOF-5. This can be associated with the higher values of the LJ interaction parameter of \ch{CH4}. For instance, at 300 K and 60 bar, the amount of \ch{H2} adsorbed in MOF-5 is approximately 0.6\%, while the amount of \ch{CH4} adsorbed is 20\%. 

\begin{figure*}[htbp]
    \centering
    \subfigure[][10 bar]{\includegraphics[scale=0.21]{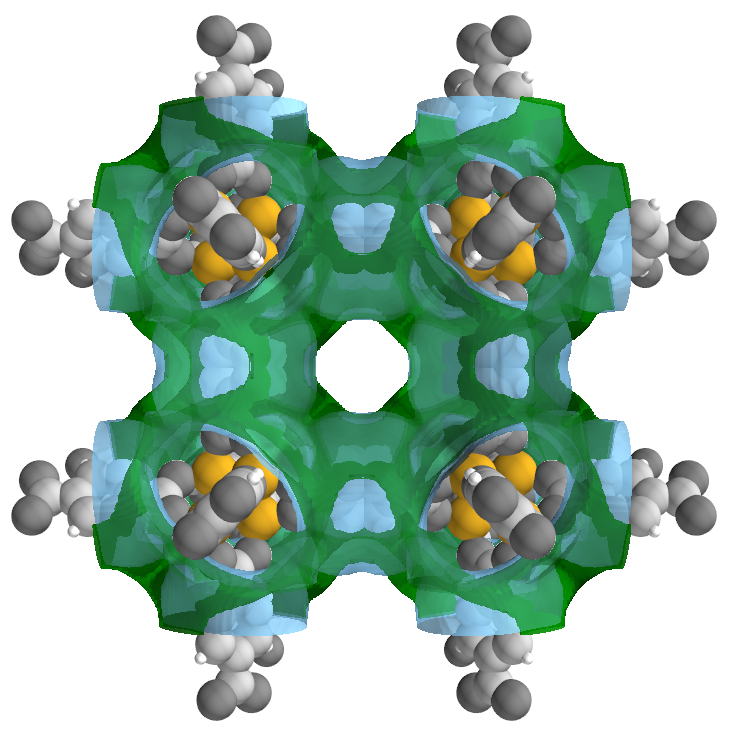}}%
    \subfigure[][30 bar]{\includegraphics[scale=0.21]{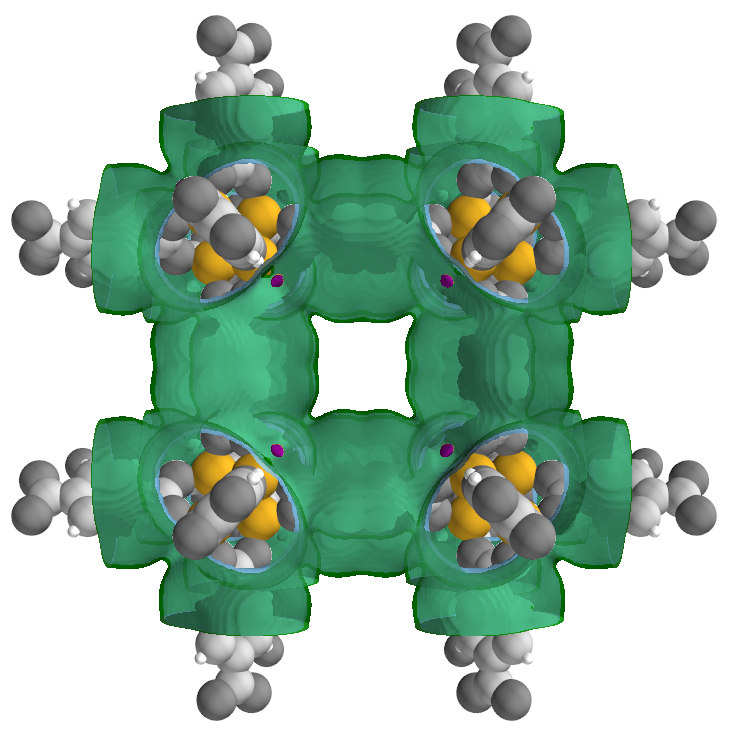}}%
    \subfigure[][60 bar]{\includegraphics[scale=0.21]{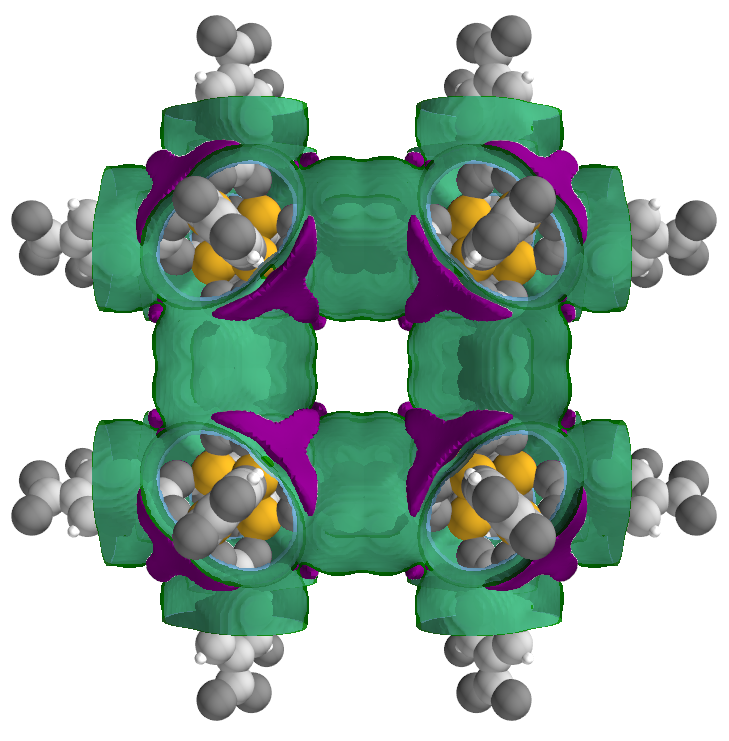}}
    \caption{Density isosurfaces of \ch{H2} in MOF-5 with local density values of 0.1 mol/L (blue), 1 mol/L (green) and 10 mol/L (purple) at the temperature of 300 K and three different pressures of (a) 10 bar, (b) 30 bar and (c) 60 bar \emph{(also see Movie S1)}. Our cDFT results presented here were calculated using $N^3 = 128^3$.}
    \label{fig:profile_H2_MOF5}
\end{figure*}

An isosurface is a surface that represents points with the same value of the field within a volume of space. Figure~\ref{fig:profile_H2_MOF5} presents the local density isosurface of \ch{H2} inside the MOF-5 at the temperature of 300 K and three different pressure values: (a) 10 bar; (b) 30 bar; and (c) 60 bar. The blue isosurface represents the region with density values of 0.1 mol/L, the green isosurface with values of 1 mol/L, and the purple isosurface with values of 10 mol/L. We can observe that for any pressure, the blue region encloses all the atoms of MOF-5, indicating the adsorption of \ch{H2} molecules around those atoms. The green isosurface has a smaller area at 10 bar but increases in the area at higher pressures, indicating an increasing concentration of \ch{H2} in the vicinity of MOF-5 atoms. Finally, the purple isosurface starts to appear at 30 bar and increases in the area at 60 bar, and appears mainly in the region around the Zn atoms. 

The concentration of \ch{H2} molecules near the Zn atoms, as outlined in Fig.~\ref{fig:profile_H2_MOF5}c, can be attributed to the stronger interaction between \ch{H2} molecules and Zn atoms, as determined by their respective LJ parameters. Additionally, the high-concentration isosurface grows due to the volume exclusion effects that arise in the vicinity of the Zn atoms when \ch{H2} is present in high concentration. The excluded volume correlations near the \ch{Zn} atoms and the organic linkers limit the physical space available to accommodate additional \ch{H2} molecules in those regions.

\begin{figure*}[htbp]
    \centering
    \subfigure[][10 bar]{\includegraphics[scale=0.21]{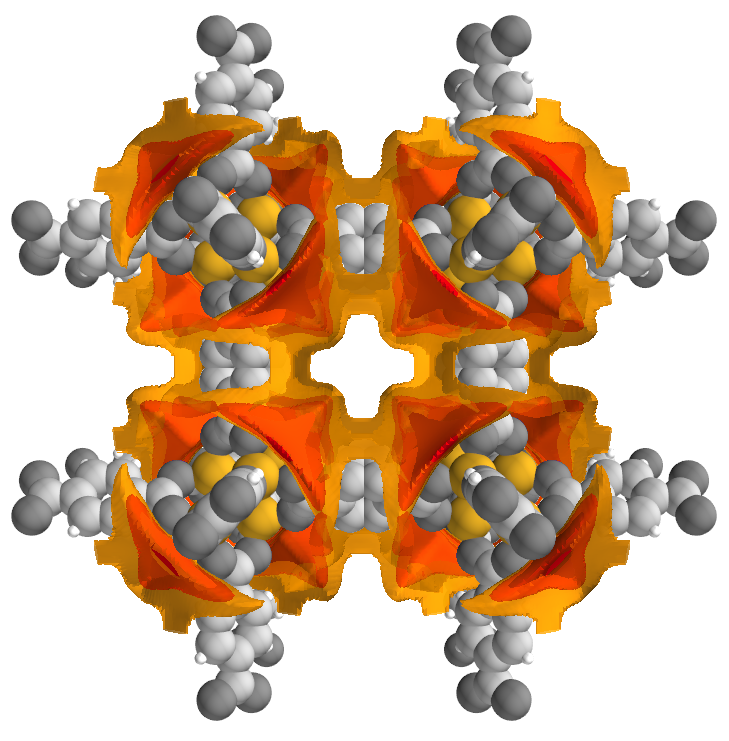}}%
    \subfigure[][30 bar]{\includegraphics[scale=0.21]{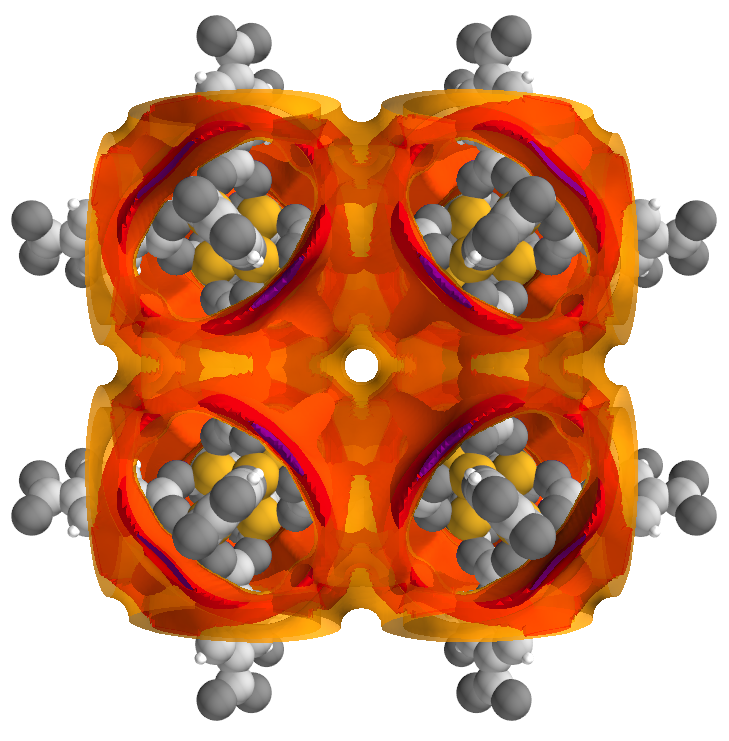}}%
    \subfigure[][60 bar]{\includegraphics[scale=0.21]{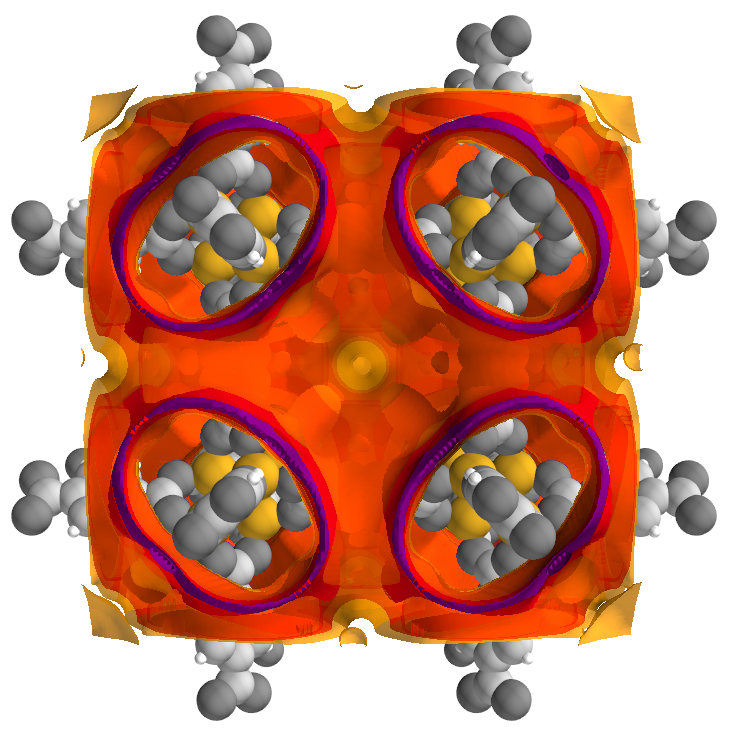}}
    \caption{Density isosurfaces of \ch{CH4} in MOF-5 with local density values of 10 mol/L (yellow), 20 mol/L (red) and 40 mol/L (purple) at the temperature of 300 K and three different pressures of (a) 10 bar, (b) 30 bar and (c) 60 bar \emph{(also see Movie S2)}. Our cDFT results presented here were calculated using $N^3 = 128^3$. }
    \label{fig:profile_CH4_MOF5}
\end{figure*} 

Figure~\ref{fig:profile_CH4_MOF5} illustrates the local density isosurfaces of \ch{CH4} within MOF-5 at 300 K and varying pressure values. The yellow isosurface represents the region with a density of 10 mol/L, the red isosurface with a density of 20 mol/L, and the purple isosurface with a density of 40 mol/L. At 10 bar, the yellow and red isosurfaces are just observed around the Zn atoms. At 30 bar, the yellow and red isosurfaces spread, and the purple isosurface begins to appear in the vicinity of the Zn atoms. At 60 bar, the yellow and red isosurfaces have spread throughout the whole box region, but the purple isosurface still surrounds the Zn atoms, indicating the highest \ch{CH4} concentration in this region.

The preferential adsorption of \ch{CH4} molecules at low and intermediate concentrations can be attributed to their strong interaction with the \ch{Zn} atoms, along with the volume exclusion effects in the proximity of these atoms, even at moderate pressures. Consequently, the concentration of \ch{CH4} molecules tends to be higher in other regions, such as in the vicinity of the organic linkers, as indicated by the negligible modification of the red isosurfaces from 30 bar to 60 bar. The highest concentration isosurface of \ch{CH4} is larger than that of \ch{H2} at 60 bar, owing to the differences in the molecular sizes of these two species.

We calculate the structure factor derived from the local density distributions to examine the spatial structure information of the adsorbed fluids within MOF-5. The structure factor $S(q)$ is calculated from Eq.~\eqref{eq:structure_fator}. Although our calculations of adsorption isotherms using a number of grid points of $N^3=64^3$ or $N^3=128^3$ lead to the same result, this is not the case for the structure factor calculations. Because $S(q)$ is obtained from the Fourier transform of $\rho(\vb*{r})$, it strongly depends on the grid size $\Delta x$ in each direction.  Therefore, the structure factors are calculated using a greater number of grid points of $N^3=128^3$. 

\begin{figure}[htbp]
    \centering
    \includegraphics[width=\linewidth]{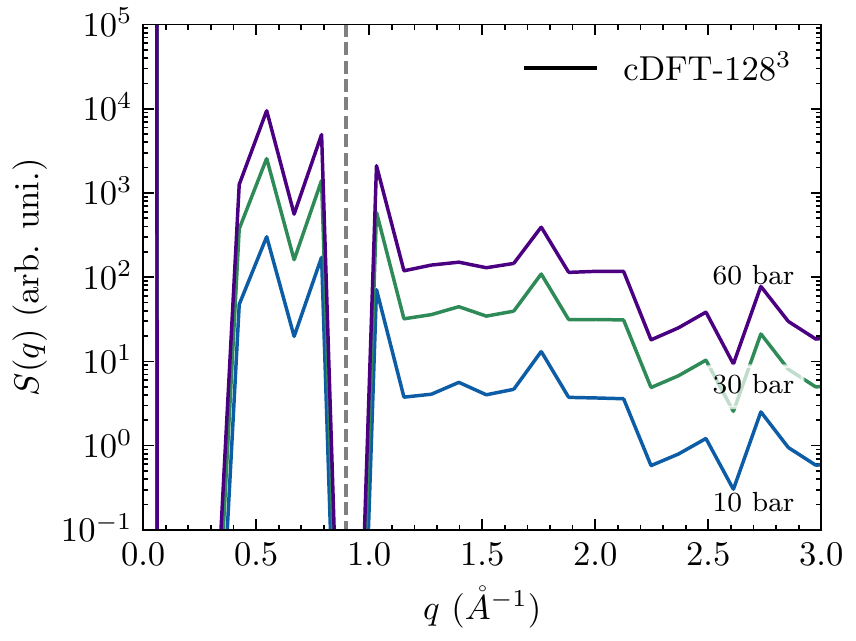}
    \caption{Structure factor of \ch{H2} in MOF-5 at 300 K and three different pressure values. Dashed lines represent specified positions discussed in the text. These cDFT results were calculated using $N^3 = 128^3$. }
    \label{fig:structure_H2_MOF5}
\end{figure}

The results presented in Figure~\ref{fig:structure_H2_MOF5} reveal the structure factor of \ch{H2} in MOF-5 at 300 K and different pressures, namely 10 bar, 30 bar, and 60 bar. Notably, the observed peaks and valleys in $S(q)$ are  correlated to the position of the MOF-5's atoms by the relation $q=2\pi/l$, with $l$ being the length scale. As example, the valley of $S(q)$ around $q = 0.9 \si{\angstrom}^{-1}$ (represented by the dashed line, characterized by $l \approx 7.0 \si{\angstrom}$) represents the radius of the cylindrical regions around the organic linkers where the \ch{H2} cannot penetrate, as illustrated in Fig.~\ref{fig:profile_H2_MOF5}. The increase in the values of $S(q)$ as we increase the pressure values is related to the compressibility of the fluid phase. Indeed, the large-wavelength limit of the structure factor gives the compressibility relation, $S(q=0) = (\pdv*{(\beta P)}{\rho_b})_T^{-1}$.
 
\begin{figure}[htbp]
    \centering
    \includegraphics[width=\linewidth]{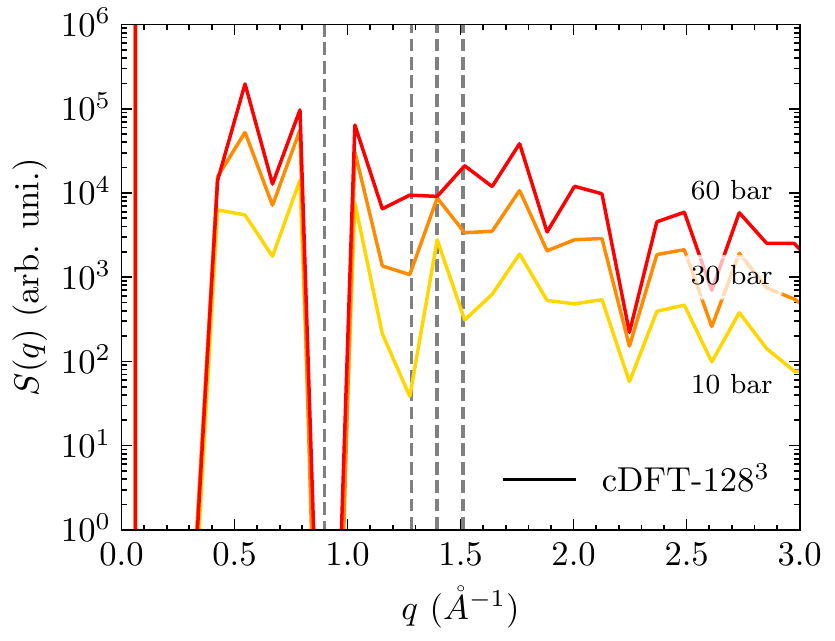}
    \caption{Structure factor of \ch{CH4} in MOF-5 at 300 K and three different pressure values. The caption is similar to Fig.~\ref{fig:structure_H2_MOF5}.}
    \label{fig:structure_CH4_MOF5}
\end{figure}

Figure~\ref{fig:structure_CH4_MOF5} presents the structure factor of \ch{CH4} fluid in MOF-5. We obtain higher values of $S(q)$ of \ch{CH4} than those values for \ch{H2}. These higher values of $S(q)$ directly correspond to a greater amount of \ch{CH4} being adsorbed within the MOF-5 structure. Again, we can note the presence of a valley of $S(q)$ around $q = 0.9\ \si{\angstrom}^{-1}$ representing the forbidden cylindrical regions for the \ch{CH4} molecules around the organic linkers, as presented in the \ch{H2} $S(q)$ curves. Moreover, we notice a significant change in the profile of $S(q)$ around $q= 1.4\ \si{\angstrom}^{-1}$ ($l \sim 4.5\ \si{\angstrom}$) at 60 bar. This change can be associated with the saturation of \ch{CH4} molecules near the Zn atoms of the MOF-5, vide Zn diameter value in Table~\ref{tab:table_solid}. The saturation effect of fluid molecules near the Zn atoms is not present in the \ch{H2} structure factor, as we can see in Fig.~\ref{fig:structure_H2_MOF5}. In addition, at a temperature of 300 K, the absolute isotherm curve for \ch{H2} exhibits a linear profile, while the absolute isotherm for \ch{CH4} demonstrates a saturation tendency at pressures approximating 60 bar. This outcome is corroborated by the results of Fig.~\ref{fig:CH4_excess_adsorption_MOF5} and \ref{fig:H2_excess_adsorption_MOF5}, which demonstrate the saturation of \ch{CH4} adsorption occurring at 80 bar, while for the \ch{H2} fluid, the saturation is observed at 300 bar. These findings are consistent with the fact that the saturation of the adsorption process takes place near the framework atoms.

In conclusion, the patterns observed in the $S(q)$ curves, as depicted in Figures~\ref{fig:structure_H2_MOF5} and~\ref{fig:structure_CH4_MOF5}, provide relevant insights into the structural behavior of the fluid within the nanoporous material. Our current cDFT calculations are limited by the number of grid points used and cannot resolve the entirety of the $S(q)$ curves. This limitation highlights the necessity of improvements on higher-resolution cDFT calculations in future works.

\section{Conclusions}
\label{sec:conclusions}

The present study highlights the importance of temperature and pressure in the gas adsorption of MOF-5 for storage applications. For MOF-5, our results indicate that \ch{CH4} adsorption is more favorable than \ch{H2} due to the higher LJ interaction parameter of \ch{CH4}. At 300 K, \ch{CH4} adsorption saturates at high pressures near the Zn atoms and tends to occupy the region near the organic linkers, while \ch{H2} concentration always increases near the Zn atoms. The results suggest that the interaction parameters, volume exclusion effects, and the size of the gas molecules influence the adsorption behavior.

By implementing of cDFT, the local density distribution of adsorbed fluids within MOF-5 can be precisely computed, enabling insights into the adsorption mechanisms of \ch{H2} and \ch{CH4} molecules in this material. Notably, the adsorption mechanism is not directly dependent on the pore description of MOF-5 as a spherical pore or its pore size. Our calculated density isosurfaces suggested that the adsorption behavior is strongly linked to the spatial distribution of the constituent atoms within the framework.

The fluid's structure factor can be derived from the local density distribution of adsorbed fluids within MOF-5. The structure factor curves can provide valuable information about the spatial distribution of fluid molecules within MOF-5. The locations and concentrations of the adsorbed molecules can be determined by comparing the peaks in the structure factor curves with the geometrical information of the MOF-5 framework.

The results presented here are significant as they provide insight into \ch{H2} and \ch{CH4} adsorption behavior in MOF-5 under different thermodynamic conditions for storage applications. The latest GCMC and cDFT works reported results at temperature values of 298 K for both fluids and 77 K for hydrogen. There is a lack of simulated data at intermediate temperatures. However, our cDFT calculations are predictive at intermediate temperatures without any additional adjustment of solid-fluid interaction parameters. Moreover, the combination of calculated fluid structure factor and small-angle scattering experiments (\emph{e.g.} SAXS and SANS)~\cite{Lee2013,Ting2015,Welborn2020} can provide a comprehensive understanding of the adsorption behavior of fluid molecules inside the MOF-5, which is essential for developing efficient gas storage and separation technologies.

\section*{Acknowledgments}

The authors wish to thank Petrobras and Shell which provided financial support through the Research, Development, and Innovation Investment Clause, in collaboration with the Brazilian National Agency of Petroleum, Natural Gas, and Biofuels (ANP, Brazil). Additionally, this research was partially funded by CNPq, CAPES, and FAPERJ.

The authors would like to thank the anonymous reviewers. Their contributions have significantly improved the quality and clarity of this publication.

\section*{Supplementary material}
The supplementary data associated with this article can be found in the online version of the Supporting Information.

We present the following files:
\begin{itemize}
    \item \emph{Supporting Information}: cDFT definitions and numerical tools.
    \item \emph{movieS1.mp4}: presents the density isosurfaces of \ch{H2} in MOF-5 with local density values of 0.1 mol/L (blue), 1 mol/L (green) and 10 mol/L (purple) at 300 K and 60 bar, with the same visualization as Fig.\ref{fig:profile_H2_MOF5}c;
    \item \emph{movieS2.mp4}: presents the density isosurfaces of \ch{CH4} in MOF-5 with local density values of 10 mol/L (yellow), 20 mol/L (red) and 40 mol/L (purple) at 300 K and 60 bar, with the same visualization as the Fig.\ref{fig:profile_CH4_MOF5}c.
\end{itemize}

\section*{The data and code availability}
The data and code that support the findings of this study are available from the corresponding author in the repository: \url{https://github.com/elvissoares/PyDFTlj}.




\bibliography{biblio.bib}

\end{document}


\title{Supporting Information to\\ Classical Density Functional Theory Reveals Structural Information of \ch{H2} and \ch{CH4} Fluids Adsorbed in MOF-5}
\author{Elvis do A. Soares}%
\email{elvis.asoares@gmail.com}
\affiliation{Engenharia de Processos Químicos e Bioquímicos (EPQB), Escola de Química, Universidade Federal do Rio de Janeiro, 21941-909, Rio de Janeiro, RJ, Brazil}%
\author{Amaro G. Barreto Jr.}%
\affiliation{Engenharia de Processos Químicos e Bioquímicos (EPQB), Escola de Química, Universidade Federal do Rio de Janeiro, 21941-909, Rio de Janeiro, RJ, Brazil}%
\author{Frederico W. Tavares}%
\email{tavares@eq.ufrj.br}
\affiliation{Engenharia de Processos Químicos e Bioquímicos (EPQB), Escola de Química, Universidade Federal do Rio de Janeiro, 21941-909, Rio de Janeiro, RJ, Brazil}%
\affiliation{Programa de Engenharia Química, COPPE, Universidade Federal do Rio de Janeiro, 21941-909, Rio de Janeiro, RJ, Brazil}%

\maketitle

\section{The Two-Yukawa representation of LJ interaction}

The attractive contribution of the LJ potential in this work is defined as
\begin{align}
    u_\text{att}(r) = \begin{cases}
        0, \quad & r<\sigma \\
        4\epsilon\left[ \left( \frac{\sigma}{r} \right)^{12}-\left( \frac{\sigma}{r} \right)^{6}\right] , & r>\sigma,
    \end{cases}
    \label{eq:attractive_contribution}
\end{align}

This attractive contribution of the LJ potential can be mapped into the sum of two Yukawa potentials as 
\begin{align}
    u_\text{ty}(r) = -\epsilon_1 \frac{e^{-\lambda_1(r/\sigma-1)}}{r/\sigma} -\epsilon_2 \frac{e^{-\lambda_2(r/\sigma-1)}}{r/\sigma},
\end{align}
and our set of parameters $\{\epsilon_i, \lambda_i\}$ was obtained by the following constraints
\begin{enumerate}
    \item LJ zero: $u_\text{ty}(\sigma) =  u_\text{lj}(\sigma) = 0$;
    \item Derivative at zero: $\left. \dv{u_\text{ty}(r)}{r}\right|_{r \to \sigma} = \left. \dv{u_\text{lj}(r)}{r}\right|_{r \to \sigma} = -\frac{24 \epsilon }{\sigma }$;
    \item Integral of potential: $\int_\sigma^\infty u_\text{ty}(r) 4 \pi r^2 \dd{r} = \int_\sigma^\infty u_\text{lj}(r) 4 \pi r^2 \dd{r} = -\frac{32}{9} \pi  \sigma ^3 \epsilon$;
    \item Integral of squared potential: $\int_\sigma^\infty [u_\text{ty}(r)]^2 4 \pi r^2 \dd{r} = \int_\sigma^\infty [u_\text{lj}(r)]^2 4 \pi r^2 \dd{r} = \frac{512}{315} \pi  \sigma ^3 \epsilon ^2$.
\end{enumerate} 
These constraints are derived from the necessary condition to accurately reproduce the thermodynamic properties of a LJ fluid. Therefore, our parameters are $\epsilon_1 = -\epsilon_2 = 1.8577 \epsilon$, $\lambda_1 = 2.5449$, and $\lambda_2 = 15.4641$. 

The set of parameters reported by \citet{Kalyuzhnyi1996}($\epsilon_1 = 2.03\epsilon$, $\epsilon_2 = 1.6438\epsilon$,  $\lambda_1 = 2.69$ and $\lambda_2 = 14.7$) and from Tang \emph{et al.}~\cite{Tang1997,Tang2001}($\epsilon_1 = -\epsilon_2 = 2.1714\epsilon$,  $\lambda_1 = 2.9637$ and $\lambda_2 = 14.0167$) are quite different from our parameters. This is due to the different constraints used and mainly the choice of the potential length in Eq.~\eqref{eq:attractive_contribution}. Here we used $\sigma$ as the potential length in the attractive potential, and the previously cited works used the hard-core Barker-Henderson diameter.

\section{Methane Fluid parameters}

In this work, we used the JZG EoS to represent the thermodynamics of LJ fluids. 

For methane molecule can be approximated by a LJ particle as reported by the TraPPE force-field~\cite{Martin1998}, with the following parameters ($\sigma_{\ch{CH4}} = 3.73$ \si{\angstrom} and $\epsilon_{\ch{CH4}}/k_B = 148.0$ K. The PC-SAFT~\cite{Gross2001} also represents the methane molecule by a LJ particle with appropriate parameters ($m_{\ch{CH4}} = 1.0000$, $\sigma_{\ch{CH4}} = 3.7039$ \si{\angstrom}, $\epsilon_{\ch{CH4}}/k_B = 150.03$ K). However, the PC-SAFT dispersive contribution was calculated using a 2-order Baker-Henderson perturbation theory.

\begin{figure}[htbp]
    \centering
    \includegraphics[width=\linewidth]{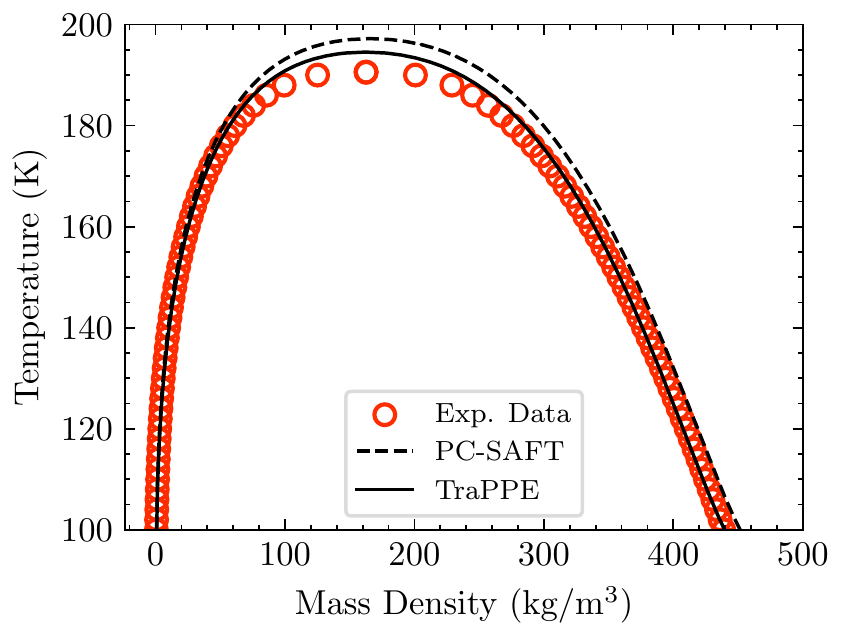}
    \caption{Vapor-Liquid Equilibria (VLE) diagram of methane fluid.  Open symbols: Experimental Data. Lines: JZG EoS with PC-SAFT parameters (\emph{dashed line}) and TraPPE parameters (\emph{solid line}). }
    \label{fig:vle_CH4}
\end{figure}

Figure~\ref{fig:vle_CH4} presents the Vapor-Liquid equibrium diagram and the vapor pressure of \ch{CH4} calculated from the JZG EoS in comparison with the experimental data. The dashed ans solid lines represent the JZG EoS with the PC-SAFT and TraPPE parameters, respectively. The VLE curve, as determined by the JZG EoS using the TraPPE parameters, shows a strong alignment with the experimental data from NIST~\cite{CH4NIST}. However, the VLE curve derived using the PC-SAFT parameters exhibits a minor discrepancy, specifically in computing the density of the liquid phase. The PC-SAFT results also overestimates the critical temperature when compared to the TraPPE results.

\begin{figure}[htbp]
    \centering
    \includegraphics[width=\linewidth]{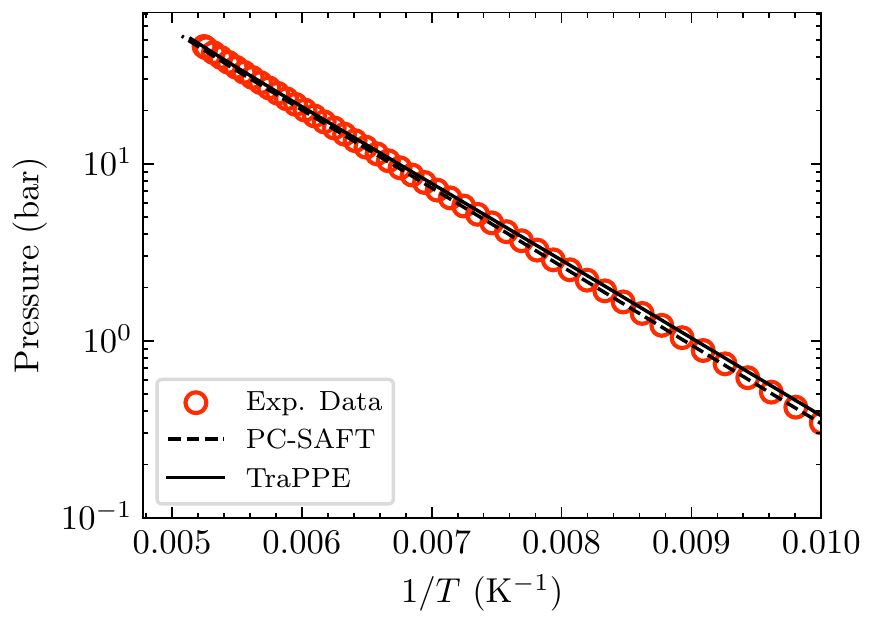}
    \caption{Vapor pressure of methane fluid. Open symbols: Experimental Data. Lines: JZG EoS with PC-SAFT parameters (\emph{dashed line}) and TraPPE parameters (\emph{solid line}). }
    \label{fig:pressure_CH4}
\end{figure}

The Figure~\ref{fig:pressure_CH4} presents the vapor saturated pressure of methane fluid. In this case, the two set of parameters can represent well the saturation pressure. 

\section{The fundamental measure theory}

The functional $F_\text{hs}$ represents the excess Helmholtz free-energy functional due to hard-sphere exclusion volume, which have been described by the modified fundamental measure theory (FMT)~\cite{Rosenfeld1989a}. In this work, we apply the White-Bear functional \cite{Yu2002a,Roth2002} for the hard-sphere Helmholtz free-energy contribution as
\begin{align}
    \beta F_\text{hs}[\rho (\vb*{r})] = \int_{\mathcal{V}} \dd{\vb*{r}} \Phi_\text{FMT}(\{ n_\alpha(\vb*{r})\}),
\end{align}
where the reduced free energy density of a mixture of hard-spheres, $\Phi_\text{FMT}$, is a function of the set of weighted densities, $n_\alpha(\vb*{r})$.

The free energy density based on the antisymmetrized version of the Rosenfeld's functional~\cite{Rosenfeld1997} is
\begin{align}
    \Phi_\text{FMT}(\{n_\alpha(\vb*{r})\}) &= \phi_1(n_3)n_0 + \phi_2(n_3)(n_1 n_2 - \vb*{n}_{v1} \cdot \vb*{n}_{v2}) \nonumber \\
    &+\phi_3(n_3)n_2^3(1 - \vb*{n}_{v2} \cdot \vb*{n}_{v2}/n_2^2)^3,
\end{align}
where the first two terms of the binomial expansion of $(1 - \vb*{n}_{v2} \cdot \vb*{n}_{v2}/n_2^2)^3$ recover the usual White Bear functional. As reported by \citet{Kessler2021} the resulting antisymmetrized functional yields accurate results for hard spheres in narrow cylindrical while retaining the full three-dimensional properties and the bulk behavior of the original White Bear functional. 

The White-Bear mark I functions are defined as
\begin{align}
    \phi_1(n_3) &=  -\ln{(1-n_3)}, \\
    \phi_2(n_3) &=  \frac{1}{1-n_3}, \\
    \phi_3(n_3) &= \frac{2n_3 +2(1-n_3)^2\ln(1-n_3)}{72 \pi n_3^2 (1-n_3)^2},
\end{align}
with the weigthed densities defined as
\begin{align}
    n_\alpha(\vb*{r}) \equiv \int_{\mathcal{V}} \dd{\vb*{r}'} \rho(\vb*{r}') \omega^{(\alpha)}(\vb*{r}-\vb*{r}'),
    \label{eq:weighted_densities}
\end{align}
whose the linearly independent weights are defined as
\begin{align}
    \omega^{(3)}(\vb*{r}) &= \Theta(d/2-|\vb*{r}|), \\
    \omega^{(2)}(\vb*{r}) &= |\grad{\Theta(d/2-|\vb*{r}|)}| = \delta(d/2-|\vb*{r}|), \\
    \vb*{\omega}^{(v2)}(\vb*{r}) &= \grad{\Theta(d/2-|\vb*{r}|)} = \frac{\vb*{r}}{r}\delta(d/2-|\vb*{r}|),     
\end{align}
and the dependent weights given by $\omega^{(0)}(\vb*{r}) = \omega^{(2)}(\vb*{r})/\pi d^2 $, $\omega^{(1)}(\vb*{r}) = \omega^{(2)}(\vb*{r})/2\pi d $, and $\vb*{\omega}^{(v1)}(\vb*{r}) = \vb*{\omega}^{(v2)}(\vb*{r})/2\pi d $. Here, $\Theta(r)$ is the Heaviside step function, $\delta(r)$ is the Dirac delta function, and $d$ is the hard-sphere diameter. We can note that 
\begin{align}
    \int_{\mathcal{V}} \omega^{(3)}(\vb*{r}) \dd{\vb*{r}} &= 4 \pi \int_0^\infty \theta(d/2-|\vb*{r}|) r^2 \dd{r}= \frac{\pi}{6}\pi d^3, \\
    \int_{\mathcal{V}} \omega^{(2)}(\vb*{r}) \dd{\vb*{r}} &= 4 \pi \int \delta(d/2-|\vb*{r}|) r^2 \dd{r} =  \pi d^2, \\
    \int_{\mathcal{V}} \omega^{(v2)}(\vb*{r}) \dd{\vb*{r}} &= 4 \pi \int \frac{\vb*{r}}{r}\delta(d/2-|\vb*{r}|) r^2 \dd{r} =  0.
\end{align} 

The Fourier transforms of these weight functions are given in the form 
\begin{align}
    \widetilde{\omega}^{(3)}(\vb*{k}) &= \int \Theta(d/2-|\vb*{r}|) e^{-i\vb*{k}\cdot \vb*{r}} \dd[3]{r} \nonumber \\
    &= \int_0^{2\pi}\dd{\phi} \int_0^{d/2}\dd{r}r^2 \int_0^{\pi}\dd{\theta} \sin{\theta}e^{-ikr\cos{\theta}} \nonumber \\
    &= \frac{2\pi}{i k} \int_0^{d/2}\dd{r}r \left[ e^{ikr}-e^{-ikr} \right] \nonumber \\
    &= \frac{\pi d^2}{k} \left[ \frac{\sin(k d/2)}{(kd/2)^2}- \frac{\cos(k d/2)}{kd/2}\right] = \frac{\pi d^2}{k}  j_1(kd/2)
    \label{eq:omega3}
\end{align}
and
\begin{align}
    \widetilde{\omega}^{(2)}(\vb*{k}) &= \int \delta(d/2-|\vb*{r}|) e^{-i\vb*{k}\cdot \vb*{r}} \dd[3]{r} \nonumber \\
    &= \int_0^{2\pi}\dd{\phi} \int_0^{\infty}\dd{r}r^2 \delta(d/2-r)\int_0^{\pi}\dd{\theta} \sin{\theta}e^{-ikr\cos{\theta}} \nonumber \\
    &= \frac{2\pi}{i k} \int_0^{\infty}\dd{r}\delta(d/2-r) r \left[ e^{ikr}-e^{-ikr} \right] \nonumber \\
    &= \pi d^2\frac{\sin(k d/2)}{kd/2} = \pi d^2 j_0(kd/2)
\end{align}
and
\begin{align}
    \widetilde{\vb*{\omega}}^{(v2)}(\vb*{k}) &= -\frac{1}{i d/2} \dv{}{\vb*{k}}\left[\tilde{\omega}^{(2)}(\vb*{k}) \right] =  - i \vb*{k} \tilde{\omega}^{(3)}(\vb*{k}) 
\end{align}

The chemical potential is obtained by the bulk composition as
\begin{align}
    &\beta \mu^{\text{hs}} = \pdv{\Phi_\text{FMT}^{(b)}}{\rho^{(b)}}.
    \label{eq:hs_chemical_potential}
\end{align}

\section{The Weigthed Density Functional Theory}
\label{app:wdft}

The term $F_\text{att}$ represents the excess Helmholtz free-energy contribution due to the particle-particle attractive interaction, which can be described by the novel weighted density functional theory (WDFT) \cite{Yu2009} as the sum of a mean-field term and a correlation contribution, respectively, in the form 
\begin{align}
    F_\text{att}[\rho (\vb*{r})] =\ & \frac{1}{2} \int_{V} \dd{\vb*{r}} \int_{V} \dd{\vb*{r}'} \rho(\vb*{r}) u_\text{att}(|\vb*{r}-\vb*{r}'|) \rho(\vb*{r}') \nonumber \\
    &+ k_B T \int_{V} \dd{\vb*{r}} \Phi_\text{corr}(\bar{\rho}(\vb*{r})),
\end{align}
where the weighted density field is given by 
\begin{align}
    \bar{\rho}(\vb*{r})= \int_{V} \dd{\vb*{r}'} \rho(\vb*{r}')\bar{\omega}(\vb*{r}-\vb*{r}')
\end{align}
where $\bar{\omega}(\vb*{r})= \Theta(d-|\vb*{r}|)/(4\pi d^3/3)$, and $\Theta(x)$ is the Heaviside function. The Fourier transform of this weight function can be calculated using Eq.~\eqref{eq:omega3} but changing $d/2$ to $d$. The correlation free-energy density is defined as
\begin{align}
    \Phi_\text{corr}(\rho) = \beta \frac{F_\text{JZG}(\rho)}{V}-\beta \frac{F_\text{hs}(\rho)}{V}-\beta \rho^2 a_\text{mft},
\end{align}
such that the excess Helmholtz free-energy of the bulk fluid coincides with the appropriated free-energy for the LJ fluid. The MFT parameter can be identified as the integral $a_\text{mft} = 2 \pi \int_0^\infty u_\text{att}(r) r^2 \dd{r} = -(16/9) \pi \epsilon \sigma^3$. 

The chemical potential is obtained by the bulk composition as
\begin{align}
    & \mu^{\text{att}} = \pdv{(F_\text{JZG}(\rho_b)/V)}{\rho^{(b)}}.
    \label{eq:att_chemical_potential}
\end{align}

\section{The Grand-Potential Functional Derivatives}

For a LJ fluid enclosed by a volume $\mathcal{V}$, with temperature $T$, and chemical potential $\mu$ specified, under the action of an external potential $\phi^\text{ext}(\vb*{r})$, the grand potential, $\Omega$, is written as
\begin{align}
    &\Omega[\rho(\vb*{r})] = F^\text{id}[\rho(\vb*{r})]  + F^\text{exc}[\rho(\vb*{r})] + \int_{\mathcal{V}} \dd{\vb*{r}} \qty(\phi^\text{ext}(\vb*{r})-\mu ) \rho(\vb*{r}) , 
\end{align}
where 
\begin{align}
    F^\text{id}[\rho (\vb*{r})] = k_B T \int_{\mathcal{V}} \dd{\vb*{r}} \rho(\vb*{r})[\ln(\rho (\vb*{r})\Lambda^3)-1],
\end{align}
and 
\begin{align}
    F^\text{exc}[\rho(\vb*{r})] =\ & F^\text{hs}[\rho(\vb*{r})] + F^\text{att}[\rho(\vb*{r})]
\end{align}

The variations are obtained keeping constant the chemical potentials $\mu$, the external potential $\phi^\text{ext}(\vb*{r})$, the temperature $T$ and the volume $V$.
\begin{align}
    \Omega[\rho + \delta \rho] =\ & F^\text{id}[\rho + \delta \rho]  + F^\text{hs}[\rho + \delta \rho] + F^\text{att}[\rho + \delta \rho] \nonumber \\
    &  + \int_{\mathcal{V}} \dd{\vb*{r}} \qty(\phi^\text{ext}(\vb*{r})-\mu) [\rho(\vb*{r})+ \delta \rho(\vb*{r})], 
\end{align}
where the ideal gas term is given by 
\begin{align}
    &F^\text{id}[\rho + \delta \rho]  = k_B T  \int_{\mathcal{V}} \dd{\vb*{r}} [\rho(\vb*{r}) + \delta \rho(\vb*{r})]\{\ln[(\rho (\vb*{r}) + \delta \rho(\vb*{r}))\Lambda^3]-1\} \nonumber \\
    &= F^\text{id}[\rho] + k_B T  \int_{\mathcal{V}} \dd{\vb*{r}} \{ \delta \rho(\vb*{r})\ln(\rho(\vb*{r})\Lambda^3) + \mathcal{O}(\delta \rho(\vb*{r})^2) \} ,
\end{align}
the HS term is given by 
\begin{align}
    &F^\text{hs}[\rho + \delta \rho]  = F^\text{hs}[\rho] \nonumber \\
    & + k_B T \sum_\alpha \iint_{\mathcal{V}} \dd{\vb*{r}} \dd{\vb*{r}'} \delta \rho(\vb*{r}) \pdv{n_\alpha(\vb*{r}')}{\rho(\vb*{r})} \pdv{ \Phi^\text{hs}}{n_\alpha(\vb*{r}')} + \mathcal{O}(\delta \rho(\vb*{r})^2) ,
\end{align}
the WDFT term is given by 
\begin{align}
    &F^\text{att}[\rho + \delta \rho]  = F^\text{corr}[\rho] \nonumber + k_B T \iint_{\mathcal{V}} \dd{\vb*{r}} \dd{\vb*{r}'} \delta \rho(\vb*{r}) \pdv{\bar{\rho}(\vb*{r}')}{\rho(\vb*{r})} \pdv{ \Phi^\text{corr}}{\bar{\rho}(\vb*{r})} \\
    &+\frac{1}{2}\iint_{\mathcal{V}} \dd{\vb*{r}}\dd{\vb*{r}'}[\rho(\vb*{r})+\delta \rho(\vb*{r})] u_\text{att}(|\vb*{r}-\vb*{r}'|)[\rho(\vb*{r}')+\delta \rho(\vb*{r}')] \nonumber \\
    =\ & F^\text{corr}[\rho] + \frac{1}{2}\iint_{\mathcal{V}} \dd{\vb*{r}}\dd{\vb*{r}'}\rho(\vb*{r}) u_\text{att}(|\vb*{r}-\vb*{r}'|)\rho(\vb*{r}') \nonumber \\
    & +k_B T \iint_{\mathcal{V}} \dd{\vb*{r}} \dd{\vb*{r}'} \delta \rho(\vb*{r}) \pdv{\bar{\rho}(\vb*{r}')}{\rho(\vb*{r})} \pdv{ \Phi^\text{corr}}{\bar{\rho}(\vb*{r})} \nonumber \\
    &+ \iint_{\mathcal{V}} \dd{\vb*{r}}\dd{\vb*{r}'}\delta \rho(\vb*{r}) u_\text{att}(|\vb*{r}-\vb*{r}'|)\rho(\vb*{r}')+ \mathcal{O}(\delta \rho(\vb*{r})^2),
\end{align} 
and the weighted densities functional derivatives being 
\begin{align}
    \pdv{n_\alpha(\vb*{r}')}{\rho(\vb*{r})} &=  \int_{\mathcal{V}} \dd{\vb*{r}''} \delta(\vb*{r}''-\vb*{r}) \omega^{(\alpha)}(\vb*{r}'-\vb*{r}'') = \omega^{(\alpha)}(\vb*{r}'-\vb*{r}) \\
    \pdv{\bar{\rho}(\vb*{r}')}{\rho(\vb*{r})} &= \int_{\mathcal{V}} \dd{\vb*{r}''} \delta(\vb*{r}''-\vb*{r}) \bar{\omega}(\vb*{r}'-\vb*{r}'') =  \bar{\omega}(\vb*{r}'-\vb*{r}) 
\end{align}

Therefore, 
\begin{widetext}
\begin{align}
    \delta \Omega &= \Omega[\rho + \delta \rho] - \Omega[\rho] \nonumber \\
    &=  k_B T  \int_{\mathcal{V}} \dd{\vb*{r}} \delta \rho(\vb*{r})\ln(\rho(\vb*{r})\Lambda^3) +  k_B T \sum_\alpha \iint_{\mathcal{V}} \dd{\vb*{r}} \dd{\vb*{r}'} \delta \rho(\vb*{r}) \omega^{(\alpha)}(\vb*{r}'-\vb*{r}) \pdv{ \Phi^\text{hs}}{n_\alpha(\vb*{r}')} \nonumber \\
    & + \iint_{\mathcal{V}} \dd{\vb*{r}}\dd{\vb*{r}'}\delta \rho(\vb*{r}) u_\text{att}(|\vb*{r}-\vb*{r}'|)\rho(\vb*{r}') + k_B T \iint_{\mathcal{V}} \dd{\vb*{r}}\dd{\vb*{r}'} \delta \rho(\vb*{r}) \bar{\omega}(\vb*{r}'-\vb*{r}) \fdv{ F^\text{corr}}{\bar{\rho}(\vb*{r})} +  \int_{\mathcal{V}} \dd{\vb*{r}} \qty(\phi^\text{ext}(\vb*{r})-\mu ) \delta \rho(\vb*{r}).
    \label{eq:delta_omega}
\end{align}
\end{widetext}

Grouping the variational terms of $\delta \rho(\vb*{r})$,we get 
\begin{widetext}
    \begin{align}
    \delta \Omega &= \int_{\mathcal{V}} \dd{\vb*{r}} \delta \rho(\vb*{r})\left[k_B T \ln(\rho(\vb*{r})\Lambda^3) + k_B T \sum_\alpha \int_{\mathcal{V}} \dd{\vb*{r}'} \omega^{(\alpha)}(\vb*{r}'-\vb*{r}) \pdv{ \Phi^\text{hs}}{n_\alpha(\vb*{r}')} + \int_{\mathcal{V}}\dd{\vb*{r}'}u_\text{att}(|\vb*{r}-\vb*{r}'|)\rho(\vb*{r}') \right. \nonumber \\
    &\left. + k_B T \int_{\mathcal{V}}\dd{\vb*{r}'} \bar{\omega}(\vb*{r}'-\vb*{r}) \fdv{ F^\text{corr}}{\bar{\rho}(\vb*{r})} + \phi^\text{ext}(\vb*{r})-\mu \right], 
    \label{eq:delta_omega_better}
    \end{align}
\end{widetext}

Finally, the functional derivative is given as follows
\begin{align}
    \fdv{\Omega}{\rho(\vb*{r})} = k_B T \ln(\rho(\vb*{r})\Lambda^3) - k_B T c^{(1)}(\vb*{r}) + \phi^\text{ext}(\vb*{r})-\mu,
\end{align}
where we define the 1st direct correlation function in the form 
\begin{align}
    c^{(1)}(\vb*{r}) =\ & -\sum_\alpha \int_{\mathcal{V}} \dd{\vb*{r}'} \omega^{(\alpha)}(\vb*{r}'-\vb*{r}) \pdv{ \Phi^\text{hs}}{n_\alpha(\vb*{r}')}\nonumber \\
    & -\beta \int_{\mathcal{V}}\dd{\vb*{r}'}u_\text{att}(|\vb*{r}-\vb*{r}'|)\rho(\vb*{r}') \nonumber \\
    &- \int_{\mathcal{V}}\dd{\vb*{r}'} \bar{\omega}(\vb*{r}'-\vb*{r}) \fdv{ F^\text{corr}}{\bar{\rho}(\vb*{r})}.
\end{align} 

The thermodynamic equilibrium condition is obtained by making
\begin{align}
    \left. \fdv{\Omega}{\rho(\vb*{r})}\right|_{\mu,V,T} &= k_B T \ln(\rho(\vb*{r})\Lambda^3) - k_B T c^{(1)}(\vb*{r}) \nonumber \\
    &+ \phi^\text{ext}(\vb*{r})-\mu= 0,
    \label{eq:equilibrium_conditions}
\end{align}
where the equilibrium value of the grand potential is a minimum with respect to the density profile. Using the definition of the ideal chemical potential such that the total chemical potential is $\mu = k_B T \ln(\rho_b\Lambda^3) + \mu_\text{exc}$, we can rewrite the Eq.~\eqref{eq:equilibrium_conditions} in the form 
\begin{align}
    \rho(\vb*{r}) = \rho_b e^{-\beta\phi^\text{ext}(\vb*{r}) + c^{(1)}(\vb*{r}) + \beta \mu_\text{exc}}.
\end{align}

\bibliography{biblio.bib}